\documentclass[a4paper,oneside,11pt,notitlepage]{report}
\usepackage{mathrsfs}
\usepackage{amssymb}
\usepackage{amsmath}
\usepackage[latin1]{inputenc}
\usepackage[dvips]{graphicx}
\usepackage[font=small,margin=\parindent, labelfont=bf]{caption}

\newcommand{\dtau}{\partial_\tau}

\newcommand{\roots}{\sqrt{s}}
\newcommand{\xbj}{{x}}

\newcommand{\xt}{\mathbf{x}_T}
\newcommand{\yt}{\mathbf{y}_T}
\newcommand{\rt}{\mathbf{r}_T}
\newcommand{\bt}{\mathbf{b}_T}
\newcommand{\pt}{{\mathbf{p}_T}}
\newcommand{\ptt}{p_T} 
\newcommand{\qt}{{\mathbf{q}_T}}
\newcommand{\kt}{{\mathbf{k}_T}}

\newcommand{\Dt}{{\mathbf{D}_T}}

\newcommand{\nabt}{\boldsymbol{\nabla}_T}
\newcommand{\itt}{\mathbf{i}_T}

\newcommand{\gtrans}{{\boldsymbol{\gamma}_T}}

\newcommand{\gt}{\gamma^0}
\newcommand{\gz}{\gamma^3}

\newcommand{\emu}{\! e_\mu \!}
\newcommand{\enu}{\! e_\nu \!}

\newcommand{\ud}{\, \mathrm{d}}

\newcommand{\intd}{\int \!}
\DeclareMathOperator{\tr}{Tr}
\newcommand{\R}{\mathrm{Re}}
\newcommand{\nc}{{N_\mathrm{c}}}

\newcommand{\half}{\frac{1}{2}}
\newcommand{\hc}{\mathrm{\ h.c.\ }}
\newcommand{\nosum}[1]{\textrm{ (no sum over } #1 )}

\newcommand{\cf}{C_\mathrm{F}}
\newcommand{\ca}{C_\mathrm{A}}
\newcommand{\df}{d_\mathrm{F}}
\newcommand{\da}{d_\mathrm{A}}
\newcommand{\nr}[1]{(\ref{#1})} 
\newcommand{\dadj}{D_{\mathrm{adj}}}
\newcommand{\ra}{R_A}

\newcommand{\gev}{\ \textrm{GeV}}
\newcommand{\fm}{\ \textrm{fm}}
\newcommand{\mb}{\ \textrm{mb}}
\newcommand{\ls}{\Lambda_\mathrm{s}}
\newcommand{\qs}{Q_\mathrm{s}}
\newcommand{\lqcd}{\Lambda_{\mathrm{QCD}}}
\newcommand{\as}{\alpha_{\mathrm{s}}}

\newcommand{\fig}{Fig.~}
\newcommand{\eq}{Eq.~}
\newcommand{\se}{Sec.~}
\newcommand{\eqs}{Eqs.~}
\newcommand{\refc}{Ref.~}
\newcommand{\refs}{Refs.~}

\begin{document}

\thispagestyle{empty}

\begin{flushright}
\textsc{HU-P-D121}\\
hep-ph/0505095
\end{flushright}


\vfill

\begin{center}

{\Large \bf  Classical chromodynamics and heavy ion collisions }
\vspace{1cm}

Tuomas Lappi\footnote{Electronic address: tuomas.lappi@helsinki.fi.}
\vspace{0.5cm}

\textit{ Theoretical Physics Division, Department of Physical Sciences and \\
Helsinki Institute of Physics\\
 P.O. Box 64, FIN-00014 University of Helsinki,
Finland}
\end{center}

\vspace{1cm}

\begin{abstract}
This paper is a slightly modified version of the introductory part of a doctoral 
dissertation  also containing the articles \mbox{hep-ph/0303076}, \mbox{hep-ph/0409328}
and \mbox{hep-ph/0409058}. 
The paper focuses on the calculation of
particle production in a relativistic heavy ion collision using the 
McLerran-Venugopalan model. The main part of the paper summarizes
the background of these numerical calculations. First we relate this
calculation of the initial stage af a heavy ion collision to our understanding
of the whole collision process. Then we discuss the saturation physics of
the small $x$ wavefunction of a hadron or a nucleus. The classical field
model of Kovner, McLerran and Weigert is then introduced before moving to 
discuss the numerical algorithms used to compute gluon and quark pair production
in this model. Finally we shortly review the results on gluon and quark-antiquark 
production obtained in the three articles mentioned above.
\end{abstract}

\vfill

\newpage
\tableofcontents

\newpage

\nocite{Lappi:2003bi,Lappi:2004sf,Gelis:2004jp}

\chapter{Introduction}
\setcounter{page}{1}
\pagenumbering{arabic}

Quantum Chromodynamics (QCD) has long since been firmly established as the correct 
theory of strong interactions; the force binding quarks into hadrons. The 
traditional phenomenological applications of weak coupling calculations have
been in processes where there is large high (transverse) momentum scale given by 
one of the scattering particles, i.e. the virtual photon in deep inelastic scattering
or a jet in $pp$ collisions. This hard scale ensures that the relevant value 
of the running coupling is small enough and higher twist corrections are suppressed.
The cross section can then be factorized  into universal parton distribution 
functions and perturbatively calculable partonic cross sections.
``Bulk'' observables, such as total multiplicities, have been considered 
so dependent on strong coupling physics that they can, even in theory, only be 
calculated from first principles by lattice calculations. 

The Relativistic Heavy Ion Collider (RHIC) at Brookhaven 
(see e.g. \cite{Adams:2005dq,Adcox:2004mh,Arsene:2004fa,Back:2004je}
for reviews of experimental results) has
been in operation since 2000. It collides heavy ions
and also lighter nuclei, deuterons and protons
at different center of mass energies up to
$\roots = 200 A \gev$. One could estimate that the production of partons with 
transverse momentum of, say, $1 \gev$ at central rapidities
probes the nuclear wavefunction at Bjorken
$\xbj$ values of $\xbj \sim 10^{-2}$. At the LHC, hopefully starting 
it operations in 2007 with $\roots= 5500 A \gev$, the corresponding 
estimate would be $\xbj  \sim 2 \cdot 10^{-4}$. The mean transverse momentum
of produced particles, mostly pions, at RHIC is $\sim 0.5 \gev$ and, as we will
discuss in \se\ref{sec:enmult}, it is not unreasonable to assume the mean
transverse momentum of the partons produced in the initial stage of the collision
to be $\sim 1 - 2 \gev$. This means that the total multiplicity and 
transverse energy at RHIC and LHC are quantities that are both dominated
by small $\xbj$ physics and involve large enough momenta to justify using weak 
coupling calculations.

Because the constituents of a nucleus are Lorentz-contracted to the same
transverse plane when boosted to high energy, one expects effects arising 
from the high density of ``wee'' partons to appear at higher values of $\xbj$ 
for larger  $A$. 
One could argue that if nonlinear effects due to the high density of partons
in the proton wavefunction start to be important at $\xbj \lesssim x_0$ (for fixed
$Q^2$), the corresponding effects for nuclei should be seen for
$\xbj \lesssim A x_0$\footnote{The scaling in the saturation model
would be  $A^{1/3\lambda}$ with $\lambda \approx 0.28$.
See e.g. \cite{Freund:2002ux} for a discussion  of the $A$-scaling in
the saturation model for small $\xbj$ deep inelastic scattering.}.
This estimate would mean that for the purpose of observing nonlinear effects
due to high parton density at e.g. transverse momenta of $1 \gev$ 
RHIC would correspond to 
$\xbj \approx 3\cdot 10^{-5}$ and the LHC to $\xbj \approx 10^{-6}$
in deep inelastic scattering on protons. Comparing to the
region accessible for HERA kinematics one could also estimate that
RHIC is like HERA with protons, while LHC will be more like
HERA for (perhaps not very heavy) nuclei.

This new range of beam energies reached at HERA, RHIC and the LHC 
is so high that ``bulk''
phenomena might become accessible to weak coupling calculations for two reasons.
Firstly, given a large enough
energy density in a large enough volume (in a nuclear collision rather than,
say, a  $pp$ experiment) the sufficiently hard scale might be given by the 
temperature of the system, viewed as a blob of \emph{quark gluon plasma}.
Another, less universally accepted, conjecture is that at
high enough energies, or equivalently small enough $\xbj$,
a sufficiently large momentum scale could be generated
by the high density of virtual ``wee'' partons in the wavefunction of
the accelerated hadron or nucleus and the nonlinear interactions of 
these partons. This latter phenomenon is referred to as \emph{saturation}.

The common feature in these two concepts is that they provide a possibility 
to analyze ``bulk'' phenomena in terms of a weak coupling constant.
One could say that RHIC has opened up a new ``firm'' region\footnote{
I owe the term ``firm'' to the talk by M. Lisa at Quark Matter 2004,
unpublished.} in the phenomenology of strong interactions,
between the soft (hadronic and stringy)
and the \emph{hard} (perturbative QCD) regions. It could be characterized
by the description of the system in terms of deconfined degrees 
of freedom, quarks and gluons, but also in terms of such high 
phase space densities of these partons that the nonlinearities
of QCD must be treated nonperturbatively. As a digression one can 
mention another interesting idea prompted by the baryon excess
at transverse momenta $\ptt \sim 3-5\gev$ observed at RHIC
\cite{Adcox:2001mf,Adler:2003kg,Adler:2003cb}. This excess over both the thermal spectra
at lower momenta and the one observed in dilute systems ($pp$ collions)
at the same momenta has been interpreted in terms of quark recombination
\cite{Fries:2004ej},
which also relies on the large phase space density of partons.


Because saturation is inherently
a high density, or strong field, phenomenon it cannot be fully understood
in terms of a \emph{perturbative} calculation (which is a power series in 
the field strength as well as the coupling constant).
However, if the saturation scale $\qs$ (the momentum scale 
characterizing the density of partons at which the nonlinearities dominate)
is large enough, one might still be able to perform a \emph{weak coupling}
calculation. The argument for a classical field approximation arises from these
circumstances; one can perform nonperturbative calculations, but use the
weak coupling to argue that quantum corrections can be neglected.
The nonlinearities of the Yang-Mills Lagrangian (see 
Appendix~\ref{sec:chromo} for the explicit form) start to dominate when
both terms of the covariant derivative $\partial_\mu + i g A_\mu$
are of equal importance. Parametrically this means that at saturation
momentum scales, $i \partial_\mu \sim \qs$, the gauge fields
involved should be of order $A_\mu \sim \qs/g$. The number
density of gluons $n$ should then be of order $n \sim A A \sim \qs^2/\as$. If the 
transverse phase space density is high enough, 
$\ud N/\ud^2\xt \sim \qs^2 / \as$ with, $\qs \gg \lqcd$, then the coupling
is weak and the occupation numbers of the quantum states involved
 $\sim 1/\as \gg 1$. This is the region 
where one expects a classical approximation to be valid.

The use of the classical field approximation in QCD is also 
interesting becauses it involves some of the most fundamental issues
in modern physics; the relation between quantum and classical theory and
the wave--particle duality. The classical field approximation is also
at present the most viable practical method to study time dependent 
phenomena in field  theory.

A classical field approximation was used to study heavy ion collisions already
in e.g. \cite{Ehtamo:1983hu}, but the idea of using the classical field 
approximation gained more traction with the model written down by McLerran and 
Venugopalan \cite{McLerran:1994ni,McLerran:1994ka,McLerran:1994vd}.

So far, however, most applications of saturation to phenomenology
have been perturbative calculations with saturation included as some kind
of phenomenological modification of perturbative gluon distributions
\cite{Kharzeev:2001gp,Kharzeev:2002pc,Kharzeev:2003wz,Kharzeev:2004if}.
While this can be  appropriate in some cases, such as pA-collisions
at RHIC where only the nucleus is in the saturation regime
\cite{Kharzeev:2002pc,Kharzeev:2002ei,Blaizot:2004wu,Blaizot:2004wv}
or heavy quark production \cite{Kharzeev:2003sk}, there are phenomena
like gluon production in central heavy ion collisions for which 
one would like to have a nonperturbative description.
It is for this reason that also classical field theory computations
with RHIC phenomenology in mind have been performed, 
starting from \cite{Krasnitz:1998ns}; they are the subject of this thesis.

Now that RHIC has already been operating for several years, the best 
that the classical field calculations have achieved are a posteriori 
explanations for what has already been observed. 
But learning from these \emph{post}dictions for RHIC we should 
be in a position to make 
more  concrete quantitative \emph{pre}dictions for the LHC results than were made
before RHIC operations started \cite{Bass:1999zq}. 
For a review on applications of classical field or saturation ideas to
RHIC phenomenology, see e.g. \cite{Lappi:2004xp,Gyulassy:2004zy,Blaizot:2004px}.

In the following we shall first try to convey a broad picture of a
relativistic heavy ion collision in Chapter~\ref{chap:rhic}.
In Chapter~\ref{chap:sat} we shall discuss saturation in the proton or
nucleus wavefunction, how it has been studied in deep inelastic 
scattering and how saturation appears in the McLerran-Venugopalan model
for the wavefunction. Then in Chapter~\ref{chap:particle}
we turn to applying these ideas to 
calculating particle production in heavy ion collisions. In 
Chapter~\ref{chap:numerics} we discuss in more detail some of
the numerical methods used in these calculations. In Chapter~\ref{chap:results}
we review the results of the three publications included in this thesis
\cite{Lappi:2003bi,Lappi:2004sf,Gelis:2004jp} before
concluding in Chapter~\ref{chap:conclusion}.

\chapter{Relativistic heavy ion experiments}
\label{chap:rhic}

\section{Spacetime picture of the collision}

We shall be interested in studying the case where two nuclei move at the
speed of light along the $x^\pm=0$-axes\footnote{See Appendix~\ref{sec:spacetime}
for the coordinate system.}.
These nuclei then collide and leave behind them, at finite
values of $\eta$ and $\tau$, some matter which is then observed in
detectors located in some region, varying between different experiments,
around $\eta=0$.
In order to properly interpret what is measured in the detectors one needs
to understand the whole collision process.

The common baseline for understanding the spacetime evolution of the matter 
formed in an ultrarelativistic heavy ion collision owes much to 
Bjorken's boost invariant hydrodynamical model \cite{Bjorken:1982qr}
(see also \cite{Blaizot:1987nc} for a spacetime picture of the
early stages of the collision). 
The Bjorken model assumes, based on experimental data from $p\bar{p}$-collisions,
that at high enough collision energies the central rapidity region, 
far enough from the fragmenting nuclei, can be described as a boost invariant
system, i.e. with particle phase space densities independent of 
spacetime rapidity $\eta.$ This model provides a relatively simple framework
for understanding the collision process, but contains several assumptions
that have to be theoretically understood and, if possible, experimentally
verified, involving several areas physics.

\begin{enumerate}
\item The initial condition at $\tau = 0$ depends on the properties
of the nuclear wavefunction at small $x$ and the dynamics of the partonic
collision process.
\item  The thermal and chemical equilibration of the matter formed
at $\tau \lesssim \tau_0$ in principle requires understanding
of time dependent, nonequilibrium Quantum Field Theory.
\item The Quark Gluon Plasma phase lasts for some fermis during
$\tau_0 \lesssim\tau \lesssim \tau_\mathrm{h}$.
For RHIC the hadronization timescale is $\tau_\mathrm{h} \lesssim 10 \fm$, 
for the LHC it is expected to be larger.
If the system reaches local thermal equilibrium, its behavior can
be described using finite temperature field theory\footnote{
For a review of recent developments in this field see
\cite{Blaizot:2003tw}.}. 
\item Finally, for $\tau  \gtrsim 10 \fm$ the system hadronizes and 
then after some time decouples and the particles fly to the detectors.
There will be no attempt to describe in detail the setup of the detectors
here, for an introduction to the detectors see e.g. \cite{Nagle:2002wj} for RHIC 
and \cite{Schukraft:2001sv,Schukraft:2001vg} for the LHC.
 \end{enumerate}

The area of interest of this thesis are the first two phases
of this picture, the initial conditions and thermalization.
In \se\ref{sec:thermal} we shall discuss the more detailed
experimental evidence for thermalization in heavy ion collisions
 and mention more fundamental
studies aimed at understanding thermalization as an aspect of time dependent
quantum field theory. Then in \se\ref{sec:enmult} we shall turn to what
can be said about these early stages of the collision by looking at 
``bulk'' observables, i.e. the transverse energy and multiplicity.

Apart from ion-ion collisions the nuclear wavefunction can also be 
independently studied in  collisions  between the nucleus and a more 
compact probe. The simplest such probe is a  lepton,
leading one to study deep inelastic scattering. We will discuss 
a saturation model approach to DIS
in \se\ref{sec:dis}. Another possibility is to use a proton or,
as is done at RHIC for technical reasons, a deuteron. Measurements
of high $\ptt$ particle spectra and jet-like correlations at RHIC
\cite{Adams:2003im,Adler:2003ii,Arsene:2004ux,Back:2003ns} 
have been essential in separating initial state nuclear or saturation 
effects, visible
already in dAu-collisions, from effects caused by the presence of 
a strongly interacting medium, i.e. final state effects\footnote{
For a theoretical calculation interpreting these measurements in 
terms of the ``color glass condensate'' see e.g. 
\refc\cite{Baier:2003hr,Kharzeev:2004bw,Kharzeev:2004yx}.}.

\section{Thermalization}
\label{sec:thermal}

The matter formed in the early stages of the collision 
is evidently so dense and strongly interacting that it
is opaque to high $\ptt$ jets  \cite{Gyulassy:2003mc}.
This can be inferred from the suppression of high $\pt$
hadrons in central AuAu-collisions \cite{Adler:2002tq,Adler:2003ii,Adams:2003im}
and the disappearence of jet-like correlations between particles
emitted in opposite azimuthal directions \cite{Adams:2003im,Adler:2004zd}.
But is this matter in thermal equilibrium?
 
The yields of different species of 
hadrons finally measured  are remarkably well reproduced by a thermal 
fit \cite{Braun-Munzinger:2003zd}
depending only on the decoupling temperature $T_\mathrm{dec}$, 
volume and baryochemical potential.
This is an indication that the hadrons are emitted from an equilibrated 
system at $T = T_\mathrm{dec}$. It has been noticed, however,
that the same kind of fit also works for $pp$ and even 
$e\bar{e}$-collisions \cite{Becattini:1995if,Becattini:1997rv,Becattini:2001fg}
and might thus result from 
from a purely combinatoric argument concerning the hadronization
process \cite{Blanchard:2004du}.

\begin{figure}[!h]
\begin{center}
\noindent
\includegraphics[width=0.7\textwidth]{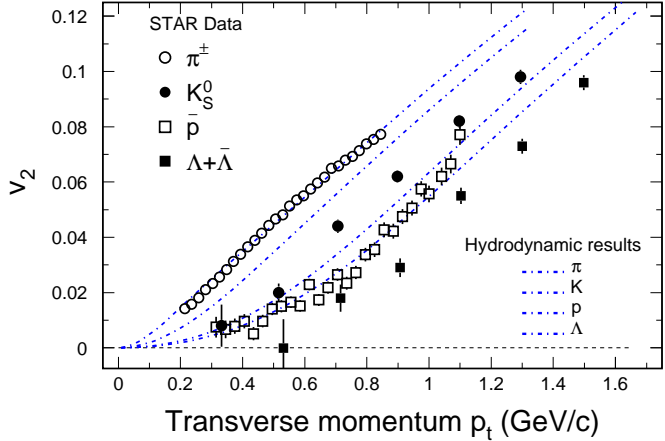}
\end{center}
\caption{ Elliptic flow $v_2$ of different particle species as a function
of trensverse momentum. Figure and experimental data
from Ref.~\protect\cite{Adams:2004bi}, hydro calculations from 
Ref.~\protect\cite{Huovinen:2001cy}.
}\label{fig:v2}
\end{figure}

\begin{figure}[!h]
\begin{center}
\noindent\includegraphics[width=0.7\textwidth]{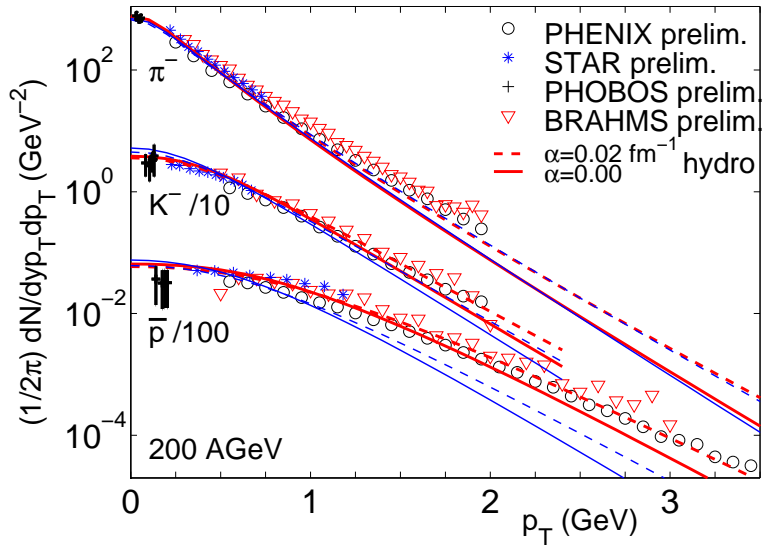}
\end{center}
\caption{ 
Right: Transverse spectra of identified hadrons compared to hydrodynamical
calculations, with $\tau_0 = 0.6 \fm$. The different theoretical curves are
for different decoupling temperatures, $T_\mathrm{dec} = 100 \gev$
for the thick lines and $T_\mathrm{dec} = 165 \gev$ thin lines. The dashed 
lines have a different initial velocity profile.
 Figure from Ref.~\protect\cite{Kolb:2002ve}.
}\label{fig:hydrospect}
\end{figure}

A perhaps more compelling argument for this scenario of thermal equilibrium and 
hydrodynamical evolution is the success of hydrodynamical models \cite{Kolb:2004pi}
in explaining a variety of ``momentum space'' experimental observables,
such as elliptic flow \cite{Kolb:2000sd} (see \fig\ref{fig:v2}) 
and particle spectra \cite{Retiere:2003kf,Eskola:2002wx} 
(see \fig\ref{fig:hydrospect}). Hydrodynamical models have been somewhat 
less successful in explaining the ``position space'' sizes
of the system as measured by HBT interferometry \cite{Magestro:2004du}
(this is referred to as the ``HBT puzzle'').

One must, of course, also be suspicious as to how much these successes
of hydrodynamical is actually due to a thermal nature of the system.
Although one is easily led to conclude from the experimental
observations that the system indeed does thermalize early enough
to justify the use of hydrodynamics, it is not yet very well 
understood theoretically what exacly is the timescale of thermalization
and how it could happen fast enough to justify the Bjorken scenario.

There is a wide literature on thermalization in  numerical calculations of real 
time  quantum field theory 
\cite{Aarts:1998td,Salle:2000hd,Salle:2000jb,Salle:2003ju,Berges:2002wr,
Berges:2002cz,Aarts:2002dj} (see \cite{Berges:2004yj} for a recent review).
These calculations, however, have so far been
mostly limited to scalar field theory (although very recently also gauge theory 
has been studied \cite{Berges:2004pu}) and to 1+1 dimensions so that they are not
applicable to RHIC physics.
One must also point out that the geometry of the initial stages of 
a heavy ion collision  is not a static 3 dimensional space, but rather one that 
is expanding in the longitudinal direction. Indeed, the phenomenologically 
most important aspect
of thermalization is not necessarily that of the thermal (exponential) 
distribution of particles in different momentum modes but the isotropization
of the momentum distribution between the transverse and the longitudinal
degrees of freedom \cite{Baier:2000sb,Arnold:2003rq}.

Most perturbative estimates, e.g. the bottom-up scenario \cite{Baier:2000sb},
generically produce quite a large thermalization time, $\tau_0 \gtrsim 3 \fm$.
Also approaches to chemical equilibration based on kinetic rate equations  
do not seem to reach chemical equilibrium fast enough to explain RHIC 
particle yields \cite{Elliott:1999uz}.
On the other hand, one could argue that if the behaviour of the
system is characterized by some quite large momentum scale, i.e. the saturation
scale $\qs \sim 1 \ldots 2 \gev$, thermalization could occur already at times
$\tau_0 \sim 1/\qs \sim 0.2 \fm$. It has been pointed out recently
(see e.g. Ref.~\cite{Arnold:2003rq,Arnold:2004ih,Arnold:2004tf,Arnold:2004ti,
Romatschke:2003ms,Romatschke:2004jh,Romatschke:2004au,Romatschke:2004ma})
 that plasma instabilities could provide the rapid
thermalization that hydrodynamical models require.

A related question is the viscosity of a strongly coupled deconfined medium.
In the weak coupling limit the shear viscosity is proportional to $1/g^4$
\footnote{For recent weak coupling calculations see e.g. 
\cite{Arnold:2000dr,Arnold:2003zc}.}. 
In ideal hydrodynamics the viscosity is neglected, so it would be important
to know what the viscosity is in the strong coupling limit. 
It has been proposed that the viscosity of $N=4$ super-Yang-Mills theory
in the strong coupling limit that can be calculated the AdS/CFT correspondence 
could give some insight into this question \cite{Policastro:2001yc}.

In this context classical field models of the nuclear wavefunction and
particle production can provide some insight into understanding
the collision process. Although the actual equilibrium state of a classical 
classical field theory is not the correct one\footnote{The 
classical theory exhibits is the famous Rayleigh-Jeans 
divergence that was one of the motivations for Planck's quantized theory
of electromagnetic radiation. The classical equilibrium state is one
where the energy is evenly distributed between all degrees of freedom, whereas
in a quantum field theory only modes with energy  $\lesssim T$ are populated.
}, one could hope that the classical theory would give a reasonable
phenomenological insight into the \emph{timescale} of thermalization.

\section{Transverse energy and multiplicity}
\label{sec:enmult}

The transverse energy and multiplicity at midrapidity are perhaps the 
most simple RHIC observables
and consequently the first ones measured.
They are also an example of the kind of quantity that cannot be calculated
in traditional collinear perturbative QCD. Phenomenological  models
traditionally used to estimate this kind of quantities have been dominated
by nonperturbative (``stringy'') physics (see e.g. \cite{Wang:1991ht}). 
The idea of gluon saturation
at small $\xbj$ opens up the fascinating possibility that such 
quantities could be understood with a weak coupling calculation based 
on the QCD Lagrangian with the nonperturbative aspects of the calculation
factorized into properties of the nuclear wavefunction.

The difference between the multiplicities and energies at
$\roots = 130 \gev$
\cite{Adcox:2001ry,Adler:2001yq,Back:2001bq,Bearden:2001xw}
and $\roots = 200 \gev$
\cite{Bearden:2001qq,Back:2002uc,Adler:2004zn,Adams:2004cb}
is not very large at the level of accuracy of the present discussion,
but let us for concreteness quote the values from \refc\cite{Adams:2004cb}
for the transverse energy in central collisions at $\roots = 200 \gev$
\begin{equation}
\frac{\ud E_T}{\ud \eta} \approx 620 \gev
\end{equation}
and the ratio of the transverse energy  to the charged multiplicity
\begin{equation}
\frac{E_T}{N_\mathrm{ch}} \approx 0.86 \gev.
\end{equation}
The particles produced in the central rapidity region are predominantly
pions (to a first approximation equal numbers of $\pi^\pm$ and $\pi^0$), 
so we can approximate that the charged multiplicity $N_\mathrm{ch}$ is 
$2/3$ of the total multiplicity, giving
\begin{equation}
\frac{\ud N_\mathrm{tot}}{\ud \eta} \approx 1100.
\end{equation}
What, then, can we infer on the properties of the system at early times
from these numbers?

In the Bjorken scenario the system undergoes an isentropic boost invariant
expansion staying in thermal equilibrium from the very early time until 
decoupling, i.e. the entropy in a unit of spacetime rapidity stays constant.
The entropy is directly proportional to the number density\footnote{
The coefficient is different for bosons and fermions and depends
on the masses of the particles, but as the measured particles are mostly 
relativistic pions (bosons with 
$\langle \ptt \rangle \approx 0.5 \gev \gtrsim m_\pi$) and the particles
in the initial state mostly gluons, we shall neglect these factors in this
discussion.}, and thus we can directly relate the measured multiplicity 
to the initial state:
\begin{equation}\label{eq:hydromulti}
\frac{\ud N^\mathrm{init}_\mathrm{tot}}{\ud \eta} \approx
\frac{\ud N^\mathrm{final}_\mathrm{tot}}{\ud \eta} \approx 1100.
\end{equation}
In the ideal hydrodynamical Bjorken expansion the energy per unit rapidity
decreases with the proper time as 
$\frac{\ud E_T}{\ud \eta} \sim \tau^{-1/3}$. Because the
volume of the unit of spacetime rapidity is $\tau$ times
the transverse area, this means that the 3-dimensional
energy density $\varepsilon$ decreases like $\tau^{-4/3}$. 
The energy decreases because an expanding system with a pressure
does $p\ud V$-work, i.e. the missing energy disappears down the beampipe to
larger rapidities. It is because of this phenomenon that the most
important aspect of thermalization for the energy and multiplicity
is the creation of a longitudinal pressure \cite{Heinz:2002rs}.
The estimate of \refc\cite{Eskola:1999fc}, assuming an early time for 
starting the decrease in the transverse energy,
is that the it could decrease even by 
a factor of $3.5$ between $\tau = 0.2 \fm$ and decoupling.
As the assumed thermalization time is very early, this could be considered
an upper limit, giving an estimate for the initial state:
\begin{equation}\label{eq:einit}
\frac{\ud E_T (\tau = 0.2 \fm)}{\ud \eta} \lesssim 2200 \gev.
\end{equation}
To express the energy density in units of $\gev/\fm^3$ that are often used
one must specify the proper time. Let us somewhat arbitrarily consider
$\tau=1\fm$ as a typical early time in the collision process when 
hydrodynamical evolution could be taken to begin. 
The decrease of a factor of $3.5$ in \refc\cite{Eskola:1999fc}
assumed a very early thermalization time. If the hydrodynamical evolution is
started only later at $\tau=1 \fm$, the transverse energy will only be able 
to decrease by a factor of $3.5 \times (0.2\fm / 1 \fm)^{1/3} \approx 2$. 
This leads to the estimate
\begin{equation}
\varepsilon (\tau = 1 \fm)
= \frac{\ud E_T(\tau = 1 \fm)}{\ud \eta} \frac{1}{\tau \pi \ra^2}
\lesssim 9 \gev/\fm^3.
\end{equation}
for the 3-dimensional energy density at $\tau=1 \fm$.
Hydrodynamical calculations with initial conditions from
different saturation models have been performed in e.g.
\cite{Eskola:2002wx,Eskola:1999fc,Eskola:2001bf,Hirano:2004rsb,Hirano:2004rs}.

Another limit may be obtained by proceeding as in e.g.
\cite{Kharzeev:2001gp,Kharzeev:2002pc,
Kharzeev:2004if,Kharzeev:2002ei,
Kharzeev:2004bw,Kharzeev:2000ph} where, analoguously to  e.g. $pp$ collisions one 
tries to directly relate the multiplicity of partons in
the initial state to the final multiplicity without assuming thermalization
or work done by a transverse pressure, simply assuming that the ratio of
the parton multiplicity in the initial state and the hadron multipliciy 
in the final state is constant.
This idea is in a sense related
to the so called \emph{parton hadron duality} \cite{Khoze:1996dn}.
If the system is not in thermal equilibrium, the entropy and thus the
particle number density is not conserved but increases\footnote{
The same happens if the matter is not an ideal fluid
but has viscosity.}. It has also been pointed out that at least in a 
perturbative calculation particle number conserving collisions thermalize
the partonic system quite slowly \cite{Mueller:1999fp} and that
particle number increasing processes are essential for thermalization
\cite{Baier:2000sb}, causing the multiplicity to increase.

If there is very little longitudinal $p \ud V$ work done by the pressure 
the energy in a unit of rapidity will be conserved:
\begin{equation}
\frac{\ud E^\mathrm{init}_T}{\ud \eta} \gtrsim 600 \gev,
\end{equation}
meaning that the energy density at $\tau=1\fm$ would be
\begin{equation}
\varepsilon(\tau = 1 \fm)
 = \frac{\ud E^\mathrm{init}_T}{\ud \eta} \frac{1}{\tau \pi \ra^2}
\gtrsim 4 \gev/\fm^3.
\end{equation}
The multiplicity, on the other hand, can increase, so our estimate
for the initial multiplicity, \eq\nr{eq:hydromulti} should be taken as an upper 
limit:
\begin{equation}\label{eq:hydromulti2}
\frac{\ud N^\mathrm{init}_\mathrm{tot}}{\ud \eta} \lesssim
\frac{\ud N^\mathrm{final}_\mathrm{tot}}{\ud \eta} \approx 1100.
\end{equation}

Essentially the difference between these two scenarios boils down to
the difference between the energy density decreasing as  $1/\tau$
for a free streaming scenario and $1/\tau^{4/3}$ for isentropic ideal
hydrodynamical expansion. If the system equilibrates early and decouples
late, the difference between $1/\tau$ vs. $1/\tau^{4/3}$ will be large.

We will return to these estimates and their meaning
in terms of the classical field model in Chapter~\ref{chap:results}.

\chapter{Parton saturation in the small $\xbj$ wavefunction}
\label{chap:sat}

\section{Parton saturation} \label{sec:sat}

The gluon distribution in a hadron or a nucleus, measured at a fixed
virtuality $Q^2$, is seen in deep inelastic scattering experiments
to grow towards smaller
$\xbj$ \cite{Breitweg:1998dz,Chekanov:2002pv}. This can be understood as
resulting from a cascade of gluons radiated from the larger $\xbj$ degrees
of freedom into the larger phase space made available by the increasing 
energy $\roots$\footnote{Or actually $\nu$, the photon energy in the target
rest frame, in DIS experiments.}.

The general idea of parton saturation 
\cite{Gribov:1984tu,Mueller:1985wy} is that as the transverse phase space
density of gluons grows large enough, it starts to be limited by gluon
fusion or, equivalently, the nonlinear interaction of the color field
with itself. More specifically, the nonabelian field strength
tensor consists of the linear part $\partial A$ and the nonlinear part
$g A^2$, and for $\partial \sim \pt \lesssim g A $ the nonlinearities
start to dominate. In this scenario the number density of gluons
would be limited from above by $n \sim AA \lesssim 1/\as.$
The momentum scale below which this happens is called 
the \emph{saturation scale} $\qs(\xbj)$.
Different authors use different conventions for the saturation scale
(we shall mention some in \se\ref{sec:othersat}), but to illustrate
the idea let us cite one definition from \cite{Mueller:1999wm}:
\begin{equation} \label{eq:muellersat}
\qs^2(x) = \frac{8 \pi^2 \as \nc}{\nc^2-1} \rho \sqrt{\ra^2-\bt^2} xG(x,\qs).
\end{equation}
This is formally an implicit equation for $\qs(x)$ but because the
r.h.s. depends on $\qs(x)$ quite weakly (logarithmically) it is easy to
get a reasonable approximation for the solution. In \eq\nr{eq:muellersat}
$\rho$ is the nuclear (baryon number) density, and together with
$2\sqrt{\ra^2-\bt^2}$, the
longitudinal extent of the nucleus at impact parameter $\bt$, it gives a 
momentum scale that is inversely proportional to the transverse area
of the nucleus. If one argues by the uncertainty principle that a 
gluon with transverse momentum $\qs$ has a transverse area $\sim 1/\qs^2$,
\eq\nr{eq:muellersat} just gives an explicit definition for the 
saturation scale as the scale where there are on the average $\sim 1/\as$
gluons overlapping in the nuclear wavefunction. When the strength of
the interaction between these $\sim 1/\as$ gluons is $\sim \as$, this
means that at the saturation scale (and lower momentum scales) the 
gluons cannot be treated as independent.

These saturation ideas have been implemented in various ways.
One straighforward application in the framework of the traditional collinear 
factorized 
formulation is to approach saturation from higher momentum scales, and include
the first nonlinear correction (the GLRMQ terms named after the authors of
\cite{Gribov:1984tu,Mueller:1985wy}) in the DGLAP evolution equations. 
This approach has been applied e.g. to charm production 
\cite{Eskola:2002yc,Eskola:2004ua}, which is an example of a quantity
that is most naturally computed in the collinear factorized framework.
The saturation argument has also been used to construct a successful model
(the Golec-Biernat \& Wusthoff model 
\cite{Golec-Biernat:1998js,Golec-Biernat:1999qd})
for describing the HERA data for deep inelastic scattering at small $\xbj$.
We shall discuss this model in \se\ref{sec:dis}.

For decreasing $x$ which, for a fixed $Q^2$, means increasing energy, 
$\qs$ is expected to grow. For large enough energy (or for large enough nuclei) 
the limit $\qs \gg \lqcd$ should be reached, making it possible to use weak 
coupling methods. 
Because the color fields are strong, the occupation numbers of quantum
states of the system are large, and it is natural to use a classical field
approximation. We shall discuss the classical field approximation in 
\se\ref{sec:mclv} before briefly reviewing some other points of view on
parton saturation in \se\ref{sec:othersat}

\section{Saturation in DIS}\label{sec:dis}

\begin{figure}[!h]
\begin{center}
\noindent
\includegraphics[width=0.7\textwidth]{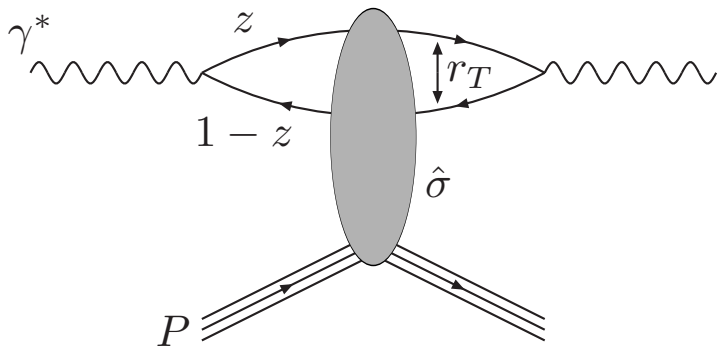}
\end{center}
\caption{In the dipole frame the incoming virtual photon splits into a 
quark-antiquark dipole of trensverse size $\rt$, which then interacts with the
target with the dipole cross section $\hat{\sigma}$.
}\label{fig:dipole}
\end{figure}

It is useful to think of deep inelastic scattering at small $\xbj$
in the dipole picture \cite{Mueller:1993rr,Mueller:1994gb,Mueller:1994jq,
Andersson:1990dp,Nikolaev:1990ja,McDermott:1999fa},
where the process is viewed as a virtual 
quark fluctuating into a color dipole, which then probes the wavefunction
of the target (see \fig\ref{fig:dipole}).

In the dipole model one factorizes the total cross section into the
probability for the virtual photon to fluctuate into a $q\bar{q}$ pair
(a color dipole) and the cross section of the dipole scattering
with the target\footnote{There are some tricky issues related to the 
Lorentz frame in which one should view the scattering process in 
the dipole model, see e.g. the discussion in
\cite{Brodsky:2002ue,Brodsky:2004hi}.}.
The total cross section can be written as 
\cite{Golec-Biernat:1998js,McDermott:1999fa}:
\begin{equation}
\sigma_{\mathrm{T,L}}(x,Q^2) = \int \ud^2 \rt \int_0^1 \ud z
|\psi_{\mathrm{T,L}}(z,\rt)|^2 \hat{\sigma}(x,Q^2,\rt). 
\end{equation}
Here the \emph{photon wave function} $\psi_{T,L}(z,\rt)$ gives the probability
for the virtual photon (T and L stand for, respectively, 
 transverse and longitudinal polarizations
of the photon) to split into a color dipole of transverse size $\rt$. The
wave function $\psi_{T,L}(z,\rt)$ includes the known QED part of the
reaction and is known analytically\footnote{
To leading order in $\alpha_{\mathrm{em}}$, which is quite sufficient
in this context.}. The exact expressions can be found in e.g. 
\refc\cite{Golec-Biernat:1998js}.

It has been shown \cite{Golec-Biernat:1998js,Golec-Biernat:1999qd} that
the HERA data on the total 
\cite{Derrick:1995ef,Aid:1996au,Derrick:1996hn,Adloff:1997mf,Breitweg:1997hz}
and the diffractive \cite{Adloff:1997sc,Breitweg:1997aa,Breitweg:1998gc}
cross sections for $x\leq 0.01$ is well reproduced by  
a saturation parametrization
\begin{equation}
\hat{\sigma}(\xbj,Q^2,\rt) = \sigma_0 (1-e^{-\rt^2 \qs(\xbj)^2/4}),
\end{equation} 
with the saturation scale  $\qs^2$ depending on $\xbj$ by
\begin{equation}
\qs^2(\xbj) = (\xbj/x_0)^{-\lambda} \gev^2.
\end{equation} 
The values found in \cite{Golec-Biernat:1998js}
were $\sigma_0= 23.03 \mb,$ 
$x_0 = 3.04 \cdot 10^{-4}$ and $\lambda = 0.288$\footnote{These are the results 
with charm quarks included. Without charm the values are somewhat different
but arguably the most important parameter in this context, $\lambda,$ 
only changes to $\lambda = 0.277$.}. 

These fits are an example of a more general hypothesis that 
the behavior of the system is controlled by a universal 
saturation scale $\qs^2(x).$ A simple demonstration of this idea
is studied in e.g. \cite{Freund:2002ux,Stasto:2000er}, where a large variety 
of small $\xbj$ deep inelastic scattering data from different experiments, 
with both protons and nuclei \cite{Adams:1995is,Adams:1994ri,Arneodo:1996kd},
 is found to follow a universal curve when plotted
as a function of the scaling variable $Q^2/\qs^2 \sim Q^2 \xbj^\lambda,$
instead of being functions of $Q^2$ and $\xbj$ separately.

\section{The classical color field of a high energy nucleus}
\label{sec:mclv}

The classical equations of motion of a nonabelian gauge field theory 
\cite{Wong:1970fu} are an interesting subject in themselves. But they are
useful especially because in some circumstances the classical field
approximation of the true quantum theory can be a very useful tool in 
understanding phenomena where high particle densities are important.
One example of such a system are the soft bosonic modes of finite temperature
field theory \cite{Blaizot:2003tw}. The example that concerns us in this
work is the small $\xbj$ wavefunction of a proton or a nucleus, where,
as we argued in \se\ref{sec:sat}, the density of gluons grows large 
(parametrically $1/\as$) and the high density gluonic system that is born 
when these quasireal gluons are freed in a relativistic heavy ion collision.

The McLerran-Venugopalan model \cite{McLerran:1994ni,McLerran:1994ka,McLerran:1994vd}
is a classical field model for the small $\xbj$
wavefunction of a hadron or a nucleus. The correspondence between 
the classical field model and a diagrammatic calculation is explored
in \cite{Kovchegov:1996ty,Kovchegov:1997pc}. Supplemented with the formulation
of a collision of two ions in \cite{Kovner:1995ja,Kovner:1995ts} 
the McLerran-Venugopalan model forms
the basis for what is referred to in this work as the classical field
model for heavy ion collisions. We shall now concentrate on the wavefunction
of one hadron or nucleus and return to colliding two of them in Chapter
\ref{chap:particle}.

The central idea behind the McLerran-Venugopalan model and one that remains at
the heart of the more general concept of the Color Glass Condensate 
is the  separation between hard (large $\xbj$) and small $\xbj$ degrees of 
freedom. The former are treated as classical sources of radiation and the latter  
as a classical color field generated by these sources. The idea is thus to have a
classical effective theory for the small $\xbj$ degrees of freedom.

Let us consider a hadron or a nucleus moving at light velocity in the positive
$z$ direction. The large $\xbj$ degrees of freedom (the valence quarks, in
a first approximation) are considered as a classical current
\begin{equation}\label{eq:source}
J^\mu = \delta^{\mu+} \delta(x^-) \rho(\xt).
\end{equation}
Let us try to justify this form somewhat. The current only has a $+$-component,
being caused by particles with a large $+$-component momentum.
The na\"ive explanation for the delta
function $\delta(x^-)$ is that when the nucleus is traveling at the speed
of light, it is Lorentz-contracted to an infinitesimal thickness. The more
proper justification is based on the Heisenberg uncertainty principle.
The large $\xbj$ degrees of freedom have a large $p^+,$ and are therefore
well localized in $x^-$. The small $\xbj$ degrees of freedom, the gluons
that form the classical field, have a smaller $p^+$ and are spread 
on a larger distance in $x^-,$ effectively seeing the sources as a delta 
function in $x^-$. The source is taken to be \emph{static} in the sense that it 
does not depend on the light cone time $x^+,$ because due to Lorentz time
dilation the evolution in $x^+$ of the source is much slower than the 
timescales of the small $\xbj$ degrees of freedom that we want to probe.

The classical color fields representing the small $\xbj$ degrees of freedom
are then computed using the Yang-Mills equations of motion
\begin{equation}\label{eq:eom}
[D_\mu,F^{\mu \nu}] = J^\nu.
\end{equation}
To consistently serve as the source term of the classical equations of
motion the current \nr{eq:source} must be covariantly conserved
\begin{equation}\label{eq:curcons}
[D_\mu,J^\mu]=0.
\end{equation}
In writing down \eq\nr{eq:source} we have implicitly assumed a gauge
where $A^-=0.$ To make the source \eq\nr{eq:source} fully gauge invariant
one would have to insert a Wilson line along the $x^+$-axis 
\cite{Jalilian-Marian:1997xn,Jalilian-Marian:2000ad} and write the 
source as
\begin{equation}\label{eq:tempwline}
J^\mu = \delta^{\mu+} \delta(x^-) W^\dag(\xt,x^+) \rho(\xt) W(\xt,x^+),
\end{equation}
where the temporal\footnote{Temporal in the sense that $x^+$ is the
\emph{light cone time}.} Wilson line is defined as
\begin{equation}\label{eq:tempwline2}
W(\xt,x^+) = P \exp \left\{ i g \int_{-\infty}^{x^+} \ud y^+ A^-(\xt,y^+) \right\}.
\end{equation}

What yet remains unknown is the transverse color charge density $\rho(\xt)$ 
in the current \nr{eq:source}. The argument of the original McLerran-Venugopalan
model \cite{McLerran:1994ni,McLerran:1994ka,McLerran:1994vd}, put on a 
firmer group theoretical footing in \cite{Jeon:2004rk}, is the following.
Let us assume that the charge density is a sum of independent color 
charges of a large number of hard partons. Then the resulting charge
density at different points of the transverse plane should be uncorrelated.
The source as a whole should be color neutral, and thus the expectation value
of the charge $\langle \rho(\xt) \rangle$ should be zero. One can also argue
by the central limit theorem that the distribution caused by a large number
of independent charges should be Gaussian. One is thus lead to a
Gaussian probablity distribution of charges
\begin{eqnarray}\label{eq:rhorho}
\langle \rho^a(\xt) \rho^b(\yt) \rangle &=& g^2 \mu^2 \delta^2(\xt-\yt)
\quad \textrm{ or }
\\
\label{eq:rhorhomom}
\langle \rho^a(\kt) \rho^b(\pt) \rangle &=& (2 \pi)^2 g^2 \mu^2 \delta^2(\kt+\pt)
\end{eqnarray}
One could also argue \cite{Lam:1999wu,Lam:2001ax} 
that the charge distribution should impose color neutrality 
at some confinement scale $\ptt \lesssim \lqcd.$ In momentum space this means
replacing the correlator \nr{eq:rhorhomom} by
\begin{equation}\label{eq:rhorhomom2}
\langle \rho^a(\kt) \rho^b(\pt) \rangle = 
(2 \pi)^2 g^2 \mu^2 \delta^2(\kt+\pt) f(\kt),
\end{equation}
where $f(\kt) \approx 0$ for $|\kt| \lesssim \lqcd$ and
$f(\kt) \approx 1 $ for $|\kt| \gtrsim \lqcd$\footnote{
In numerical simulations this kind of a modification is essential to control the 
Coulombic ``tails'' of the classical fields when studying finite size nuclei
\cite{Lappi:2003bi,Krasnitz:2002ng,Krasnitz:2002mn}.}.

One can now solve the equations of motion \nr{eq:eom}. The solution is most 
easily found in the covariant gauge $\partial_\mu A_\mathrm{cov}^\mu=0.$
One can find a solution with only one component of the gauge field is 
nonzero, namely $A_\mathrm{cov}^+(\xt,x^-).$ In this case \eq\nr{eq:eom} becomes 
a 2-dimensional Poisson equation
\begin{equation}
-\nabt^2 A_\mathrm{cov}^+ = \delta(x^-)\rho(\xt).
\end{equation}
We can formally write the solution as
\begin{equation}\label{eq:apluscov}
A_\mathrm{cov}^+ = - \delta(x^-)\rho(\xt) /  \nabt^2.
\end{equation}
Note that there is an infrared singularity in \eq\nr{eq:apluscov}. The most
natural prescription to solve this amibiguity is to impose the constraint 
$\int \ud^2 \xt \rho(\xt)=0,$ i.e. to require that the source as a whole
is color neutral. Imposing color neutrality at a shorter length scale will
also remove this ambiguity.

The covariant gauge solution has the advantage of being localized on the light 
cone in the $t,z$-plane, but its interpretation in terms of partons is not
very clear. To interpret the classical field in terms of 
quasi-real Weizs\"acker-Williams gluons we must transform the field into 
the light cone (LC) gauge. This gauge transformation can be done
 using the path ordered exponential
\begin{equation} \label{eq:pathorder}
U(\xt,x^-) = \mathrm{P} \exp \left\{ ig \int_{-\infty}^{x^-} \ud y^-
A_\mathrm{cov}^+(\xt,y^-)
\right\},
\end{equation}
giving
\begin{eqnarray} \label{eq:LCsol}
A^\pm_\mathrm{LC} & = &  0 \\
A^i_\mathrm{LC} & = & \frac{i}{g} U(\xt,x^-) \partial_i U^\dag(\xt,x^-).
\end{eqnarray}
The light cone gauge solution is not localized on the $x^+$-axis, unlike the
one in covariant gauge. Instead, for $x^- >0$ it is a transverse pure gauge field.
The field strength tensor $F^{\mu \nu},$ however, is nonzero only on the light
cone $x^-=0$.

The problem of understanging the small $\xbj$ wavefunction of the hadron or 
nucleus has now been reduced to a simple model depending on one phenomenological
parameter describing charge density of the hard sources, $\mu$, the gauge
coupling $g$, which, in the classical field approximation, is just a constant,
and the geometrical properties of the hadron or nucleus. One can try to estimate
the value of $\mu$ from the parton distribution functions \cite{Gyulassy:1997vt}
or just treat it as a parameter to be determined from experiment.

The source charge density $\mu$ increases as one probes smaller values
of $\xbj,$ because the number of partons with higher $\xbj$ than the value
of interest are counted in the classical source. 
To study this one needs a more refined description of the longitudinal structure of 
the source \cite{Lam:2000nz}. One can then develop a renormalization group 
equation to see how the hard source develops as gluons of smaller 
and smaller $\xbj$ are integrated out of the classical fields and
included in the source. This renormalization group equation is the
so called JIMWLK equation
\cite{Jalilian-Marian:1997xn,Jalilian-Marian:1997jx,Jalilian-Marian:1997gr,
Jalilian-Marian:1997dw,Weigert:2000gi,Iancu:2001ad,Mueller:2001uk,
Iancu:2001md,Iancu:2002aq,
Iancu:2002tr,Blaizot:2002xy,Rummukainen:2003ns,Iancu:2002xk,Iancu:2003xm}.

\paragraph{Relation to the dipole model}
In the classical field approximation the dipole 
cross section can be expressed in terms
of Wilson lines in the background field of a nucleus
\cite{Balitsky:1995ub,Buchmuller:1996xw,Buchmuller:1998jv,Mueller:2001fv}
\begin{equation}
\hat{\sigma}(\rt) = 
\frac{1}{\nc}
\int \ud^2 \bt \tr 
\left\langle 1 - U^\dag(\bt + \rt /2 ) U(\bt- \rt / 2) \right\rangle
\end{equation}
where $U(\xt)$  in the Wilson line in the fundamental representation:
\begin{equation}
U(\xt) = \mathrm{P} \exp \left\{ ig \int_{-\infty}^{\infty} \ud x^-
A_\mathrm{cov}^+(\xt,x^-)
\right\}.
\end{equation}

The correlators of the Wilson lines in the McLerran-Venugopalan model
were studied numerically in \cite{Gavai:1996vu} and analytically 
in \cite{Kovchegov:1999yj}. For a discussion of the different
representations and conventions on the saturation scale see also
\cite{Lam:2002mg}. \refc\cite{Kovner:2000pt}
discusses the relation between different approaches to JIMWLK equation
and how the same equation arises when considering the quantum
evolution as a property of either the target wavefunction or the (dipole)
probe.

By calculating the gluon distribution in the McLerran-Venugopalan
model one can relate the strength of the color source to the saturation
scale $\qs$ defined by A. Mueller  and Yu. Kovchegov
(see e.g. Ref.~\cite{Krasnitz:2002mn,Kovchegov:2000hz})
\begin{equation}\label{eq:satscale}
\qs^2 = \frac{g^4 \mu^2 \ca }{4 \pi}
\ln \left( \frac{g^4 \mu^2}{\lqcd^2} \right).
\end{equation}

One must also note that there is a dependence, although only a logarithmic
one, on an infrared cutoff in Eq. \nr{eq:satscale}. In analytical calculations
one often argues that the relevant cutoff is $\lqcd$, because at that scale
confinement physics sets in. In numerical calculations the implicit
infrared cutoff is the system size $\pi \ra^2$.

\section{Other views of saturation}\label{sec:othersat}

Let us then briefly mention some other views and aspects of parton
saturation before proceeding to calculate gluon production in 
the classical field model in Chapter~\ref{chap:particle}.

An interesting proposal, discussed in e.g. in 
\refc\cite{Armesto:1996kt,Nardi:1998qb,Satz:2002qj,Digal:2002bm,Digal:2003sg},
is to relate parton saturation to percolation. As we have noted
one can argue by the Heisenberg uncertainty principle that an individual
parton with transverse momentum $\ptt$ occupies an area $\sim 1/ \ptt^2$ 
in the transverse plane. Saturation then corresponds to the situation
where the wavefunctions of these partons begin to overlap and, 
because of their strong interactions, behave collectively. With this
picture in mind one can consider saturation as a percolation phase 
transition. One would then expect to see clear signals of critical behavior
in the system, say as a function of centrality in a heavy ion collision.

The EKRT-model \cite{Eskola:1999fc} is a
final state saturation model, where the saturation scale $p_\mathrm{sat}$
is determined from a geometrical saturation condition for the gluons 
freed in a heavy ion collision. The same scale $p_\mathrm{sat}$ serves
as an infrared cutoff for the pQCD calculation giving the number of 
produced gluons in terms of usual parton distribution functions. 
Although the value of $p_\mathrm{sat}$ is close to $\qs$, the saturation
scale in the nuclear wavefunction, it is not possible to give an explicit
formula relating the two due to the different concepts of initial and
final state saturation.
 
The saturation scale, or radius, $\qs = 1/R_\mathrm{s}$ in the work of
Golec-Biernat and Wusthoff \cite{Golec-Biernat:1998js} 
or Rummukainen and Weigert \cite{Rummukainen:2003ns} is the same
as the saturation scale $\qs$ of Kovchegov et. al. 
except with $\ca$ replaced by $\cf$.
The technical reason for this is that they consider correlators
of Wilson lines in the fundamental representation, whereas the
gluon distribution that Mueller and Kovchegov consider involves
the same correlator in the adjoint representation.
For a comment on the relation to the saturation scale used by
E. Iancu et. al., see Ref.~\cite{Lam:2002mg}.

We are using the old notation of of Krasnitz et. al. \cite{Krasnitz:1998ns}.
In their newer work \cite{Krasnitz:2002mn} they use the parameter $\ls$
defined as $\ls \equiv g^2 \mu$.

\chapter{Particle production in the classical field model}
\label{chap:particle} 

\section{The classical field model}\label{sec:kmclw}

\begin{figure}[!h]
\begin{center}
\noindent
\includegraphics[width=0.8\textwidth]{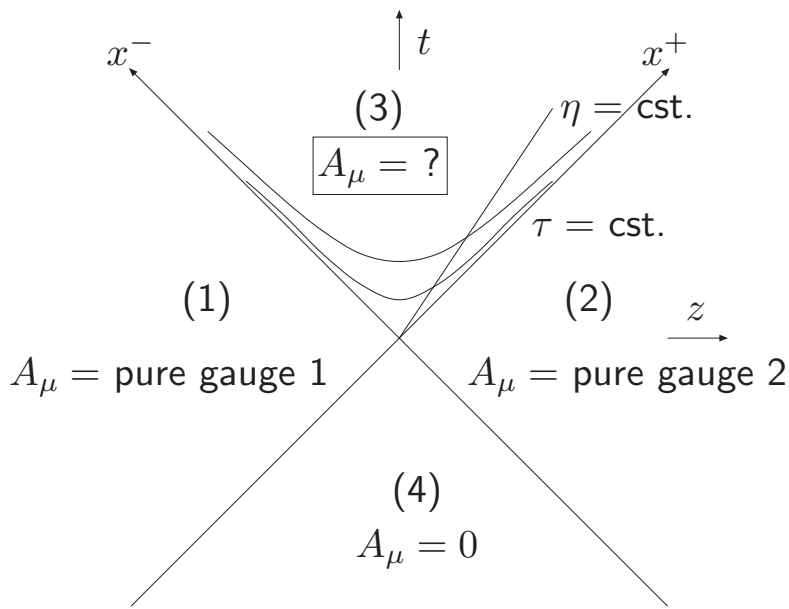}
\end{center}
\caption{Spacetime structure of the KMcLW model. The gauge field in regions
(1) and (2) are pure gauge fields on one nucleus. They are used to find the initial 
condition to find the gauge field in the region (3).
}\label{fig:spacet}
\end{figure}

Let us then turn, following the approach of Kovner, McLerran and Weigert
(KMcLW) \cite{Kovner:1995ts}, to studying the collision between two nuclei in the 
McLerran-Venugopalan model. We want to calculate the  classical gauge 
field in the forward light cone using as initial condition the fields 
of the two nuclei, given by \eq\nr{eq:LCsol}. This calculation in the
light cone gauge was formulated and performed perturbatively
to leading nontrivial order in \cite{Kovner:1995ja,Kovner:1995ts,Gyulassy:1997vt}
and to leading order on one of the sources and all orders in the other in
\cite{Dumitru:2001ux}. The same computation can also be done in the covariant gauge
\cite{Kovchegov:1997ke,Kovchegov:1998bi,Kovchegov:2000hz}.

We start with a static current of the two nuclei,
\begin{equation}\label{eq:twonucl}
 J^{\mu} =  \delta^{\mu +}\rho_{(1)}(\xt)\delta(x^-) 
+ \delta^{\mu -}\rho_{(2)}(\xt)\delta(x^+),
\end{equation}
anticipating the fact that we will be working in a gauge where this form
is covariantly conserved.
In the regions $x^->0, \ x^+ < 0$ (1) and $x^+ >0, \ x^- < 0$ (2)
(see \fig\ref{fig:spacet}) the field is given by the pure gauges
as in \eq\nr{eq:LCsol}
\begin{equation} \label{eq:apureg}
A_{(m)}^i = \frac{i}{g} e^{ i\Lambda_{(m)}} \partial_i 
e^{ -i\Lambda_{(m)}}, \textrm{ with }
\nabt^2 \Lambda_{(m)}(\xt) = -g \rho_{(m)}(\xt), \quad m=1,2.
\end{equation}
Here we have introduced the notation $\Lambda_{(m)}$ for the solution of the 
Poisson equation. The quantity $\Lambda_{(m)}$ is related to to the covariant
gauge fields of the nuclei by $gA^+_\mathrm{cov} = \delta(x^-)\Lambda_{(1)},
\quad gA^-_\mathrm{cov} = \delta(x^+)\Lambda_{(2)}$.

We then choose to work in a temporal gauge 
$A_\tau= (x^+A^- + x^-A^+) / \tau =0$\footnote{See Appendix~\ref{sec:spacetime}
for our conventions concerning the coordinate system.}. This will enable us to 
use a Hamiltonian formalism. It also matches smoothly to the conditions 
$A^-=0$ for $x^-=0$ and $A^+=0$ for $x^+=0$ that, as discussed in
\se\ref{sec:mclv}, enable us to drop the
temporal Wilson lines from the current, \eq\nr{eq:tempwline}
and use the simple form \nr{eq:twonucl}. 
In this gauge the remaining components of the gauge field 
are the transverse components $A^i$ and the ``longitudinal'' 
component $A_\eta= -\tau^2 A^\eta = x^+A^- - x^- A^+$.

Inside the future light cone (region (3) in \fig\ref{fig:spacet}) the gauge 
fields satisfy the equations of motion in vacuum. What is needed is the initial
condition for solving these equations. These initial conditions can be obtained
by requiring that the fields in the different regions match smoothly on the 
light cone. In practice the initial conditions can be obtained by inserting
the ansatz
\begin{eqnarray} \label{eq:iniansatz}
A^i &=& \theta(-x^+)\theta(x^-) A^i_{(1)}
+ \theta(x^+)\theta(-x^-) A^i_{(2)}
+\theta(x^+)\theta(x^-) A^i_{(3)} \\
A^\eta &=&  \theta(x^+)\theta(x^-) A^\eta_{(3)}
\end{eqnarray}
into the equation of motion \nr{eq:eom} and requiring that the singular terms
arising from the derivatives of the $\theta$-functions cancel. 
In this way one gets the following initial conditions
for the gauge field in the future light cone:
\begin{eqnarray}\label{eq:trinitcond}
A^i_{(3)}|_{\tau=0} &=& A^i_{(1)} + A^i_{(2)} \\
\label{eq:longinitcond}
A^\eta_{(3)}|_{\tau=0} &=& \frac{ig}{2}[A^i_{(1)},A^i_{(2)}].
\end{eqnarray}
The equations of motion with these initial conditions can then be solved 
either numerically or perturbatively in the weak field limit. Let us
first look at the equations of motion in more detail and then 
review the weak field solution of \cite{Kovner:1995ts} in
\se\ref{sec:pertsol}.

\section{Boost invariant Hamiltonian}
\label{sec:contham}

The initial conditions, \eqs\nr{eq:trinitcond} and \nr{eq:longinitcond},
 are boost invariant. Therefore it is
natural to assume that also the solution of the equations of motion will be
independent of rapidity. To maintain a consistent boost invariance of the
solution one must not perform gauge transformations that depend on the
rapidity $\eta.$ This limitation reduces the longitudinal gauge field
$A_\eta$ to an adjoint scalar field, and we will denote it by $A_\eta \equiv \phi.$
We could equally well choose $A^\eta=-A_\eta/\tau^2$ as our canonical variable;
 this would simply cause powers of $\tau$ to appear in different locations
 (see Appendix~\ref{sec:spacetime}).
The appearence of an adjoint scalar is analoguous to the way the time component of 
the gauge field
becomes an adjoint scalar when dimensionally reducing high temperature
gauge theory to a 3 dimensional effective theory \cite{Blaizot:2003tw}.
Gauge field theory with an adjoint scalar field in 
the $\tau, \eta$-coordinate system has been extensively studied in
\cite{Makhlin:1996dt,Makhlin:1996dr,Makhlin:1998zi,Makhlin:2000nw,
Makhlin:2000nx, Makhlin:2000ny,Makhlin:2000xf}.

In the gauge $A_\tau=0$  and with the assumption of boost invariance
we end up with the following action of a 2+1-dimensional gauge theory 
with a scalar field in the adjoint representation,
\begin{equation}\label{eq:contaction4}
S = \int \ud \eta \ud^2 \xt \ud \tau  \ \tau \tr \bigg\{
\dot{A_i}\dot{A_i}
- \frac{1}{2} F_{ij} F_{ij}
+ \frac{1}{\tau^2} \dot{\phi}^2
- \frac{1}{\tau^2}  [D_i,\phi] [D_i,\phi]
\bigg\}.
\end{equation}
The fields in this action are now all functions of the proper time $\tau$ and the
transverse coordinate $\xt$. A dot denotes a derivative with respect to $\tau$, 
i.e. $\dot{A}_i \equiv \dtau A_i$.
The explicit time dependence in the action
is caused by the $\tau, \eta$-coordinate system. The
expanding longitudinal geometry of the coordinate system reflects the physical
situation; the collision region is expanding with light velocity as the
two sources move apart from each other.

In the $A_\tau=0$ gauge we can go to the Hamiltonian formulation easily. 
Defining the canonical momenta (electric fields)
\begin{eqnarray}
E^{ia} &\equiv& \frac{\delta S}{\delta\dot{A_i^a}} = \tau \dot{A^a_i} 
= - \tau \dot{A^{ai}}
\\
\pi^a &\equiv& \frac{\delta S}{\delta\dot{\phi^a}} = \frac{1}{\tau}\dot{\phi^a}
\end{eqnarray}
we get the Hamiltonian density \cite{Makhlin:1996dr}
\begin{eqnarray}\label{eq:contham}
\mathcal{H} &=& 2 \tr [ E^i \dot{A_i} + \pi \dot{\phi}  ] - \mathcal{L}
\\ \nonumber
&=&
\tr \bigg\{  \frac{1}{\tau} E^iE^i + 
\frac{\tau}{2}F_{ij}F_{ij} 
+\tau \pi^2+ \frac{1}{\tau} [D_i,\phi] [D_i,\phi]
\bigg\}.
\end{eqnarray}
The first term in the Hamiltonian is the kinetic energy term of the 
transverse electric fields. The second term is the potential energy of the transverse
gauge fields, which, in a 3-dimensional language, would be called the 
 $z$-component of the magnetic field. In 2+1-dimensional gauge theory 
the magnetic field only has this one component.
The third and fourth terms are the kinetic and potential energy,
or electric and magnetic terms, of the scalar field.

The Hamiltonian equations of motion in the vacuum can be obtained in 
the standard way:
\begin{eqnarray}\label{eq:conteqmo1}
\dot{A}_i &=& t^a \frac{\delta H}{\delta E^i_a} = \frac{1}{\tau} E^i
\\ \label{eq:conteqmo2}
\dot{\phi} &=& t^a \frac{\delta H}{\delta p^a} = \tau \pi 
\\ \label{eq:conteqmo3}
\dot{E}^i &=& - t^a \frac{\delta H}{\delta A_i^a} =
\tau[D_k,F_{ki}] - \frac{ig}{\tau}[\phi,[D_i,\phi]]
\\ \label{eq:conteqmo4}
\dot{\pi} &=& - t^a \frac{\delta H}{\delta \phi^a} 
= \frac{1}{\tau} [D_i,[D_i,\phi]].
\end{eqnarray}

\paragraph{Initial conditions for the Hamiltonian variables}

The initial condition for the transverse gauge fields is given directly by 
the sum of the two transverse pure gauges, \eq\nr{eq:trinitcond}.
Because the transverse electric field is proportional
to $\tau$: $E^i = \tau \dot{A}_i$ its initial condition is simply
$E^i(\tau=0,\xt ) = 0$. The initial condition for $A^\eta$
is given by the commutator of the two transverse pure gauges,
\eq\nr{eq:longinitcond}. Because $\phi \equiv A_\eta =-\tau^2 A^\eta$,
this means that $\phi(\tau=0,\xt)=0$. The corresponding momentum, on the
other hand, is $\pi = \dot{\phi}/\tau = -2 A^{\eta} -\tau \dot{A}^\eta$ and
thus the initial condition for $\pi$ is 
$\pi(\tau=0,\xt) =  -i g \left[ A^i_{(1)},A^i_{(2)} \right]$.

\section{Gluon production in the weak field limit}
\label{sec:pertsol}

Let us then calculate the number of gluons produced in a collision of two
ultrarelativistic nuclei to the leading nontrivial order in the classical sources.
Our calculation is essentially the same as performed in \cite{Kovner:1995ts},
exept that we will use the notation of the previous section.
We will expand in powers of the dimensionless functions $\Lambda_{(m)}(\xt)$
defined in \eq\nr{eq:apureg}. They are proportional to the source strengths,
$\Lambda_{(m)}(\xt) \sim g \rho_{(m)}(\xt) \sim g^2 \mu$, so the dimensionless
parameter that must be small in this computation is $g^2 \mu/\kt$.
We will have to expand the fields to 
order $\Lambda^2$ and thus the multiplicity, which is quadratic in the fields,
will be proportional to $\Lambda^4$.

Before solving the equations let us first discuss the definition of the
multiplicity. Our starting point is the requirement that the energy
can be expressed as an integral over momentum modes:
\begin{equation}\label{eq:hmodes}
H = \int \ud^2 \kt n(\kt) |\kt|.
\end{equation}
This defines the differetial multiplicity $ n(\kt) $, assuming a massless
dispersion relation  $E_\kt = |\kt|$ and a given partition of the energy
into momentum modes. We shall also take advantage of the 
equipartition of energy between the coordinates and the momenta in a classical 
mechanical system and only use only the momentum part of the Hamiltonian 
from \eq\nr{eq:contham},
\begin{eqnarray}\label{eq:hequi}
H &\approx& 2 \int \ud^2 \xt
\tr \left[  \frac{1}{\tau} E^iE^i +\tau \pi^2 \right]
\\ \nonumber
&=& 2 \int \frac{\ud^2 \kt}{(2 \pi)^2}
\tr \left[  \frac{1}{\tau} E^i(\kt)E^i(-\kt) +\tau \pi(\kt) \pi(-\kt) \right],
\end{eqnarray}
to define the multiplicity of gluons.
The ``$\approx$'' should be taken as an equality in a time averaged sense. 
The advantage of using only the momenta is that we can easily express the 
kinetic part of the Hamiltonian in terms of the Fourier transforms of the momenta.
Equating the expressions \nr{eq:hmodes} and \nr{eq:hequi} we then find
the differential multiplicity
\begin{equation}\label{eq:defmulti}
n(\kt) = \frac{1}{(2 \pi)^2}\frac{2}{|\kt|} \left[
\frac{1}{\tau} E^i(\kt)E^i(-\kt) +\tau \pi(\kt) \pi(-\kt)
\right].
\end{equation}
Unlike the energy, the multiplicity as defined in \eq\nr{eq:defmulti} is not a gauge
invariant concept. The prescription we shall use is to fix the Coulomb gauge
in the transverse plane, $\partial_i A_i =0$.

The initial conditions for the canonical variables are, expanded to order
$\Lambda^2$, 
\begin{eqnarray} \label{eq:ainitpert}
A_i(0,\xt) &=& -\frac{1}{g}\left(\partial_i \Lambda_{(1)}(\xt)+
\partial_i \Lambda_{(2)}(\xt) \right)
+ \frac{i}{2g}\left[\partial_i \Lambda_{(1)}(\xt), \Lambda_{(1)}(\xt)\right]
\nonumber \\ &&
+\frac{i}{2g}\left[\partial_i \Lambda_{(2)}(\xt), \Lambda_{(2)}(\xt)\right]
\\ \label{eq:piinitpert}
\pi(0, \xt) &=& \frac{i}{g} \left[\partial_i \Lambda_{(1)}(\xt),
 \partial_i \Lambda_{(2)}(\xt)\right].
\end{eqnarray}
It is easiest to fix the Coulomb gauge already in the initial condition. This 
removes the lowest order terms in $A_i(\tau=0,\xt)$, giving 
\begin{multline}
A_i^\mathrm{Coul}(0, \xt) = \frac{i}{2g} \left(\delta_{ij}-
\frac{\partial_i \partial_j}{\nabt^2} \right)
\bigg( \left[ \Lambda_{(1)}(\xt),\partial_j \Lambda_{(2)}(\xt)\right]
\\
 +\left[ \Lambda_{(2)}(\xt),\partial_j \Lambda_{(1)}(\xt)\right]
 \bigg).
\end{multline} 
From here on we will drop the superscript ``Coul'' and consider all fields in the
Coulomb gauge for the rest of this section.
Now that both $A_i$ and $\pi$ are of order $\Lambda^2$, it is easy to linearize the
equations of motion \eqs\nr{eq:conteqmo1}--\nr{eq:conteqmo4}.
They can then be solved by Fourier transforming with respect
to the transverse coordinate
\begin{eqnarray} 
\dot{E}^i &=& \partial_\tau(\tau \dot{A}_i) = \tau \nabt^2 A_i
\\
& \Longrightarrow & \left(\tau^2 \partial_\tau^2 + \tau \partial_\tau
+ \tau^2 \kt^2 \right)A_i(\tau,\kt) =0
\\
\dot{\pi} &=& \partial_\tau \left( \frac{1}{\tau} \dot{\phi}\right)
= \frac{1}{\tau} \nabt^2 \phi
\\
& \Longrightarrow&\left(\tau^2 \partial_\tau^2 -\tau \partial_\tau
+ \tau^2 \kt^2 \right)\phi(\tau,\kt) =0.
\end{eqnarray}
The solutions of these equations are Bessel functions
\begin{eqnarray}
A_i(\tau,\kt) & =& A_i(0,\kt)J_0(|\kt| \tau)
\\
\phi(\tau,\kt) &=& \frac{\tau}{|\kt|} \pi(0,\kt) J_1( |\kt| \tau).
\end{eqnarray}

Using the asymptotic expansions of the Bessel functions
we get the expectation value of the multiplicity
\begin{multline} \label{eq:oscmulti}
\langle n(\tau,\kt) \rangle
= \frac{1}{(2 \pi)^2}\frac{2}{\pi \kt^2 } \bigg\{
\kt^2 \sin^2\left(|\kt|\tau -\frac{\pi}{4} \right) 
\langle A_i^a(0,\kt) A_i^a(0,-\kt)\rangle
\\
+ \sin^2\left(|\kt|\tau -\frac{3\pi}{4} \right)
\langle \pi^a(0,\kt) \pi^a(0,-\kt)\rangle
\bigg\}
\end{multline}
The correlators $\langle A_i^a(0,\kt) A_i^a(0,-\kt)\rangle$ and 
$\langle \pi^a(0,\kt) \pi^a(0,-\kt)\rangle$ can be calculated using the initial
conditions, \eqs\nr{eq:ainitpert} and \nr{eq:piinitpert}, the Poisson equation
\nr{eq:apureg} relating the $\Lambda$'s to the charge density $\rho$
and the correlator of the charge densities, \eq\nr{eq:rhorho}. One obtains
\begin{eqnarray}
\langle A_i^a(0,\kt) A_i^a(0,-\kt)\rangle &=&
\pi \ra^2 \frac{\nc(\nc^2-1)}{g^2} (g^2 \mu)^4
 \int \frac{\ud^2\pt \ud^2 \qt}{(2\pi)^2}
 \\ \nonumber &&
\delta^2 (\kt -\pt-\qt) 
 \frac{\pt^2 \qt^2 -(\pt\cdot \qt)^2}{\kt^2 \qt^4 \pt^4}
\\
\langle \pi^a(0,\kt) \pi^a(0,-\kt)\rangle &=&
\pi \ra^2 \frac{\nc(\nc^2-1)}{g^2} (g^2 \mu)^4
\int \frac{\ud^2\pt \ud^2 \qt}{(2\pi)^2}
\\ \nonumber && 
\delta^2 (\kt -\pt-\qt) 
\frac{(\pt\cdot \qt)^2}{\qt^4 \pt^4}
\end{eqnarray}
Because we only used the momenta to define the multiplicity, we are left with
the oscillating $\sin^2 |\kt| \tau$ factor in \eq\nr{eq:oscmulti}. We will replace
this oscillating factor with its time average of $1/2$, which is equivalent to 
using also the contribution from the fields. To be more explicit; when we 
argued by the equipartition of energy and chose to use only the momenta,
we effectively replaced $1 = \sin^2 |\kt| \tau + \cos^2 |\kt| \tau $
by $2 \sin^2 |\kt| \tau$. If we now use the time average 
$\overline{\sin^2 \left( |\kt| \tau \right) } = 1/2$ 
the result will be equivalent to using both
the fields and the momenta.

With this prescription the $(\pt\cdot \qt)^2$-terms cancel between the transverse
and scalar fields and we end up with the infrared divergent result
\begin{equation}\label{eq:irdivpert}
\langle n(\kt) \rangle =   \frac{\pi \ra^2}{(2 \pi)^2}
\frac{\nc (\nc^2-1) g^6 \mu^4}{\kt^2} \frac{1}{\pi}
\int \frac{\ud^2 \pt}{(2 \pi)^2} \frac{1}{\pt^2 (\kt-\pt)^2}.
\end{equation}
This is the result found in \cite{Kovner:1995ts} (some  small errors of
\cite{Kovner:1995ts} were corrected in \cite{Gyulassy:1997vt}, where the 
constant factors are correct). This result is analoguous to the bremsstrahlung
result of Gunion and Bertsch \cite{Gunion:1981qs} but with a different,
in our case infrared divergent,
form factor for the source (see discussion in \cite{Gyulassy:1997vt}).

If we regulate the integral \eq\nr{eq:irdivpert} with a mass 
$m$, replacing $\pt^2$ and $(\kt-\pt)^2$ with $\pt^2 + m^2$ and 
$(\kt-\pt)^2+m^2$,  it can be evaluated and we obtain
\begin{equation}\label{eq:regpert}
\langle n(\kt) \rangle =   \frac{\pi \ra^2}{(2 \pi)^3}
\frac{1}{\pi} \frac{\nc (\nc^2-1) g^6 \mu^4}{\kt^4} 
\ln \frac{\kt^2}{m^2}. 
\end{equation}
This result still has an unintegrable singularity at $\kt = 0$ and therefore 
does not give a finite result for the total multiplicity. It nevertheless has 
some important properties that will also hold for the full numerically computed 
result. The result is proportional to the transverse area of one nucleus,
$\pi \ra^2$. If we assume that the infrared divergence if the integral 
\nr{eq:irdivpert} should be regulated at the scale $m \sim g^2 \mu$, where,
as discussed previously, our perturbative expansion breaks down, the 
multiplicity can be written in a general form as 
\begin{equation}\label{eq:multgeneral}
\langle n(\kt) \rangle = \frac{\pi \ra^2}{g^2} f(\kt/g^2 \mu).
\end{equation}
The numerical calculation \cite{Lappi:2003bi} shows that when all orders 
in the source are included in the calculation the integrated multiplicity is 
indeed finite. The total multiplicity is then given in the form
\begin{equation}\label{eq:pertfn}
\frac{\ud N}{\ud \eta} = \int \ud^2 \kt \langle n(\kt) \rangle 
= \frac{\pi \ra^2}{g^2} (g^2 \mu)^2 f_N,
\end{equation}
where 
\begin{equation}
f_N \equiv \int \ud^2 \left(\frac{\kt}{g^2 \mu}\right) f\left(\frac{\kt}{g^2 \mu}\right)
\end{equation}
is a numerical constant (i.e. independent of $g,\pi \ra^2, \textrm{ and } \mu$).
By the same kind of argument the transverse energy should be given by
\begin{equation}\label{eq:pertfe}
\frac{\ud E_T}{\ud \eta} = \int \ud^2 \kt \langle n(\kt) \rangle |\kt| 
= \frac{\pi \ra^2}{g^2} (g^2 \mu)^3 f_E.
\end{equation}
Numerically computing the constants $f_N$ and $f_E$ is the purpose of 
the first article in this thesis \cite{Lappi:2003bi}. We will discuss
the numerical method for this calculation, developed in \cite{Krasnitz:1998ns}, 
in Chapter~\ref{chap:numerics} and turn now to quark pair production.

\section{Fermion pair production}

Given the classical fields corresponding to gluon production
a natural question to ask is: how are quark antiquark pairs produced by
these color fields? Formally quark production is suppressed by
$\alpha_\textrm{s}$ and group theory factors compared to gluons, so
in a first approximation we should be able to treat the quarks as a small
perturbation and neglect their backreaction on the color fields
(see e.g. \cite{Eskola:1996ce} for a comparison of quark antiquark
pair and gluon production in a pQCD framework).
Heavy quark production  is calculable already in perturbation
theory and in the weak field limit quark pair production from
the classical field model reduces to a known result in $k_\mathrm{T}$-factorized
perturbation theory \cite{Gelis:2003vh}.
It is, however, unknown how much the full inclusion of the
strong color fields changes the result.
Understanding light quark production would address the question of
chemical equilibration; turning the color glass condensate into a
\emph{quark} gluon plasma. There have been calculations of quark 
production in heavy ion collisions from instantons \cite{Shuryak:2002qz},
 from static ``stringy'' longitudinal electric fields
using the Schwinger mechanism \cite{Kajantie:1985jh,Gatoff:1987uf,
Kluger:1992gb,Cooper:1993hw} and in the pQCD + saturation model 
\cite{Eskola:1999fc,Eskola:1996ce}. The approach of calculating
the fermion Green's functions in a space-time dependent background
gauge field in \refs\cite{Dietrich:2003qf,Dietrich:2004eb} is more similar
to our approach in \refc\cite{Gelis:2004jp}, but has not been directly
applied to the gauge fields from the classical field model discussed
in \se\ref{sec:kmclw}.

\begin{figure}[!h]
\begin{center}
\noindent
\includegraphics[width=0.8\textwidth]{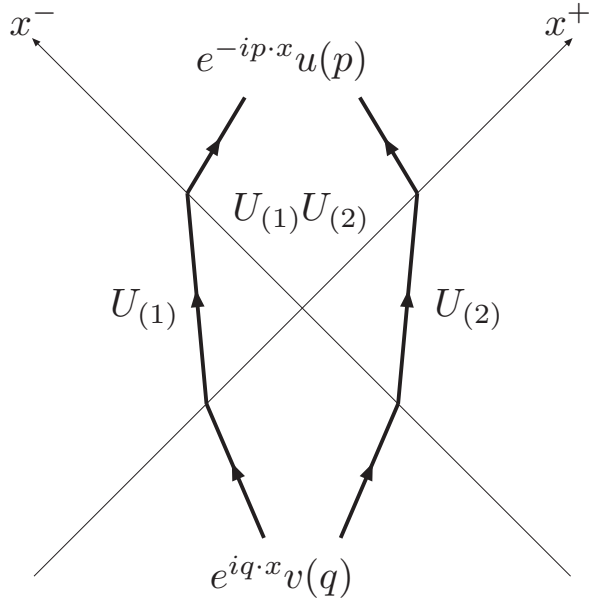}
\end{center}
\caption{The quark antiquark pair production calculation in spacetime. We start at
$x^+<0, \ x^- < 0$ with a 
negative energy spinor on shell, $e^{i q \cdot x} v(q)$. When the antiquark
line meets the first color source, it is kicked off mass shell by the 
color field of the source in the form
of the Wilson line $U_{(1)}$ or $U_{(2)}$. When colliding with the 
second color source
the antiquark gets another kick. The label $U_{(1)}U_{(2)}$ refers to the 
Abelian case, in which the field in the future light cone is a pure gauge field
given by the product $U_{(1)}U_{(2)}$. In the non-Abelian case the field 
is known only numerically. In the end the wave function is projected
to positive energy states $e^{-ip\cdot x}u(p)$ to find the pair production 
amplitude.
}\label{fig:qqbarspacet}
\end{figure}

The calculation of quark pair production, outlined in more detail
in Ref.~\cite{Gelis:2004jp}, proceeds by solving the Dirac equation in the
background color field of the two nuclei. This can be done analytically
for the Abelian theory. The QED calculation, of interest for electron
positron pair production in ultraperipheral collisions, is done e.g. in 
\refc\cite{Baltz:1998zb}. 

Let us briefly summarize the theoretical background of the calculation
developed in \refc\cite{Baltz:2001dp}. 
It is well known that the \emph{cross section} to produce \emph{one} quark antiquark
pair can be calculated by evaluating the Feynman, or time ordered, propagator for 
the spinor field. The time ordered propagator is also the quantity that is
usually obtained when 
computing Feynman diagrams, because the Wick theorem applies to time ordered
propagators. The main observation of \refc\cite{Baltz:2001dp} was that
the \emph{expectation value of the number of quark antiquark pairs}. 
is related to the \emph{retarded} quark propagator. Because the Wick theorem
does not apply to the retarded propagator it is difficult to calculate 
in perturbation theory. However, because calculating the retarded propagator
is equivalent to solving the Dirac equation as an initial value problem, 
it is easy to formulate as a numerical calculation. One has to 
solve the Dirac equation with a negative energy plane wave as the initial 
condition. The resulting spinor is then projected to positive energy states
at long enough times after the collision to obtain the quark antiquark pair
multiplicity.

One can give a physical interpretation for this calculation in terms of the
Dirac hole theory, where the vacuum is interpreted as a Dirac sea with all the 
negative energy quark states filled. Pair creation by an external field is then
interpreted as the gauge field giving enough energy to a negative energy quark
to lift it to a positive energy state, leaving a ``hole'' in the Dirac sea
of negative energy quarks. This hole is then interpreted as an antiquark. The 
Dirac sea interpretation should of course be considered as a pedagogical 
explanation. The real justification for this approach is the detailed 
calculation in \refc\cite{Baltz:2001dp}, where one starts by expressing
the expectation value of the number of quark antiquark pairs in terms of 
creation and annihilation operators and then, by lengthy manipulations, 
relates this expression to the retarded propagator.

The initial condition for $t \to -\infty$ is a negative energy plane wave.
Similarly to the QED case, one can find analytically the solution for
the regions $x^\pm>0, \ x^\mp<0$, because the gauge fields in these regions 
are pure gauges. These then give the initial condition
at $\tau=0$ for a numerical solution of the Dirac equation in the
forward light cone $\tau>0$. To find the number of quark pairs
one then projects the numerically calculated wave function to a positive
energy plane wave at some sufficiently large time $\tau$.
The structure of the calculation in different regions in spacetime is demonstrated
in \fig\ref{fig:qqbarspacet}.

A major technical challenge in this calculation is the coordinate system.
In order to include the hard sources of the color fields, the colliding nuclei,
only in the initial condition of the numerical calculation, one wants
to use the proper time $\tau$ instead of the Minkowski time $t$. Unlike
the gluon production case, where one was able to assume strict boost invariance,
one now has a nontrivial correlation between the rapidities of the quark and
the antiquark. Thus, although the background gauge field is boost invariant,
one must solve the Dirac equation 3+1 dimensions.

A crucial point in the calculation is the 
choice of the longitudinal variable to be used with $\tau$. Possible
choices are $\eta,z,\textrm{ or }x^\pm$.  To obtain the correct result we must have
a dimensionful longitudinal variable, such as $z$ or $x^\pm$ to
parametrize the $\tau=0$ surface.  This is because one must be able to
represent longitudinal momentum scales in coordinate space. For $\tau > 0$ the
corresponding longitudinal coordinate could be constructed as $\tau
e^{-\eta},$ but for $\tau=0$ this is not possible. To enable a
symmetric treatment of both branches in \fig\ref{fig:qqbarspacet}, 
one should choose $z$ as the longitudinal variable, with $x^\pm =
(\sqrt{\tau^2+z^2}\pm z)/\sqrt2=(|z|\pm z)/\sqrt{2}$ at $\tau=0.$

One is thus led to solve numerically the Dirac equation in a curvilinear
coordinate system. The covariant formulation of spinors in a curved
space is a fascinating subject in itself \cite{Birrell:1982ix}, but as
in this context one is merely decribing ordinary Minkowski spacetime
with curved coordinates it was ultimately found in \refc\cite{Gelis:2004jp}
that the most practical solution for this particular case is to 
simply change the coordinates without transforming the spinors themselves.
The Dirac equation thus obtained is naturally linear in the fermion field
$\psi,$ but also has an unpleasant property. 
Due to the choice of coordinate system and the decision to use the flat 
\emph{vierbein} the coefficients of the time and space derivatives 
in the resulting Dirac equation, \eq\nr{eq:diractauz}, depend on the coordinates.
(The vierbein determines the relation between the flat tangent space
where the Dirac matrices are defined to the point in spacetime.
See \cite{Birrell:1982ix} for the theory or Appendix A of \refc\cite{Gelis:2004jp}
for a brief explanation.)
As will be explained in \se\ref{sec:implicit} this makes finding a stable
discretization scheme harder, and one has to use an implicit discretization scheme.

\chapter{Details of the numerical calculation}
\label{chap:numerics}

\section{Classical chromodynamics on the lattice}
\label{sec:ccd}

Let us first review, following \cite{Krasnitz:1998ns}, 
how the 2+1-dimensional Hamiltonian developed in \se\ref{sec:contham}
can be cast in a form suitable for numerical simulation on a transverse
lattice with a  finite lattice spacing. This lattice version of the
calculation has been used in many numerical studies, e.g.
\cite{Lappi:2003bi,Krasnitz:1998ns,Krasnitz:2000gz,
Krasnitz:1999wc,Krasnitz:2001qu,Krasnitz:2003jw,Jalilian-Marian:2003mf}.

Wilson's lattice gauge field theory was first cast in the Hamiltonian form 
by Kogut and Susskind \cite{Kogut:1974ag,Kogut:1982ds}. The classical
Hamiltonian formulation has been used extensively in e.g. calculations
of the sphaleron transition rate in electroweak theory
\cite{Krasnitz:1995xi,Ambjorn:1995xm,Ambjorn:1997jz,Bodeker:1999gx}.
Our notations for lattice gauge theory, formulated in terms of 
link matrices and plaquettes, are explained in Appendix~\ref{sec:wilson}.

We will start from the Wilson action (with real Minkowski time, because
we want to obtain the equations of motion in real time). Then we will take
the continuum limit in the $t$ and $z$-directions, leaving only the 
transverse coordinate discretized. This will allow us to change the coordinates
to $\tau,\eta$ and restrict ourselves to $\eta$-independent field configurations.
Finally we will discretize the new time coordinate $\tau$ using the leapfrog 
algorithm.

The Wilson action for gauge fields is
\begin{equation} \label{eq:mw}
S = - \frac{2\nc}{g^2} \sum_x \left[
\sum_{1 \leq i < j \leq 3} \left( 1-\frac{1}{\nc}\R \ \tr U_{i,j}(x) \right)
- \sum_{i=1}^3 \left( 1-\frac{1}{\nc}\R \ \tr U_{0,i}(x) \right)
\right],
\end{equation}
where the $-$ sign between the two terms is needed because the fields are in 
real, and not imaginary time.  The coordinate $x$ stands for 
the different points in the 3+1-dimensional spacetime lattice.
We then take the continuous limit in the $t$ and 
$z$-directions to obtain
\begin{equation}\label{eq:puolicont}
S = \intd \ud z \ud t \sum_{\xt} \Bigg[
-\frac{2\nc}{g^2} \left( 1-\frac{1}{\nc}\R \tr U_{1,2} \right)  
+ \frac{2}{g^2} \sum_{i=1}^2 \R \tr [M_{0i}-M_{3i}]
+ \tr F_{03}^2
\Bigg],
\end{equation}
where we have set the lattice spacing $a$ to unity.
Here we have defined the partially continuous limit (continuous 
in the $\mu$-direction, $\mu=0,3$ 
but discrete in the $i$-direction, $i=1,2$) of the plaquette
\begin{multline}\label{eq:puolicontb}
M_{\mu i}(\xt)  =  \frac{g^2}{2}\left[A_\mu^2(\xt)+A_\mu^2(\xt+\itt)\right] 
-ig A_\mu(\xt)(\partial_\mu U_i(\xt))U^\dag_i(\xt) 
\\
- ig  A_i(\xt+\itt) (\partial_\mu U^\dag_i(\xt))U_i(\xt)
- \frac{1}{2} (\partial^2_\mu U^\dag_i(\xt))U_i(\xt) 
\\
-g^2 A_\mu(\xt) U_i(\xt) A_\mu(\xt+\itt) U^\dag_i(\xt)
\nosum{i},
\end{multline}
where the $t,z$ arguments have been left out for brevity and
$\itt$ denotes a unit vector in the $i$-direction, $i=1,2$.

To enable a Hamiltonian formulation we must then choose a temporal 
gauge condition $A_\tau=0.$ We shall also assume that the field
configurations are boost invariant, i.e. independent of $\eta$.
To consistently impose this boost invariance we must also forbid
$\eta$-dependent gauge transformations, which reduces
the longitudinal gauge field $A_\eta$ into a scalar field transforming
in the adjoint representation of the gauge group. We shall consequently
denote $A_\eta = \phi$. With these restrictions the action becomes
\begin{multline}\label{eq:puolicont2}
S = \int \ud  \eta \ud \tau \sum_{\xt} \tau \Bigg\{
\frac{1}{g^2} \sum_i  \tr \left(\dot{U_i}^\dag  \dot{U_i}\right)
-\frac{2\nc}{g^2} \left( 1-\frac{1}{\nc}\R \ \tr U_\bot \right)
\\
+ \frac{1}{\tau^2} \tr \dot{\phi}^2
- \frac{1}{\tau^2} \sum_i  \tr 
\Big( \phi -  \tilde{\phi}_i \Big)^2
\Bigg\},
\end{multline}
where $\dot{U}_i$ denotes the derivative with respect to $\tau$. 
We have also introduced the parallel transported scalar field
\begin{equation}\label{eq:fitilde}
\tilde{\phi}_i(\xt) \equiv U_i(\xt)\phi(\xt+\itt)U^\dag _i(\xt) \nosum{i}.
\end{equation}
The difference $\phi -  \tilde{\phi}_i$ is a \emph{covariant}
difference, because the link matrices $U_i$ have been used to 
parallel transport $\phi(\xt + \itt)$ so that it gauge transforms
at the point $\xt$, i.e. as
\begin{multline}
\tilde{\phi}_i(\xt) \Longrightarrow \left[ V(\xt) U_i(\xt) V^\dag(\xt+\itt) \right] 
\left[V(\xt+\itt)\phi(\xt+\itt) V^\dag(\xt+\itt)\right] \times
\\
\left[ V(\xt+\itt) U^\dag _i(\xt) V^\dag(\xt)\right]
= V(\xt)\tilde{\phi}_i(\xt) V^\dag(\xt)
\end{multline}
and can thus be compared with $\phi(\xt)$ in a gauge invariant way. 

To perform the Legendre transformation from the Lagrangian to the 
Hamiltonian formalism we need to replace the time derivatives
in the action \nr{eq:puolicont2} by the canonical momenta, the 
left invariant transverse electric fields 
\begin{equation}\label{eq:Efield}
E^i_a = \frac{2i\tau}{g^2} \tr t_a U_i \dot{U_i}^\dag  \nosum{i}
\end{equation}
and the longitudinal electric fields
\begin{equation}
\pi = \dot{\phi}/\tau.
\end{equation}
In terms of these canonical variables the Hamiltonian density becomes
\begin{equation}\label{eq:kogutsusskind}
\mathcal{H} =  \frac{g^2}{\tau}\tr E^iE^i +
\frac{2\nc\tau}{g^2} \left( 1-\frac{1}{\nc}\R \tr U_{1,2} \right)
\\ +
\tau \tr \pi^2 +
\frac{1}{\tau} \sum_i
\tr \left(  \phi - \tilde{\phi}_i \right)^2.
\end{equation}

The dynamics of the theory is defined by the Hamiltonian 
\nr{eq:kogutsusskind} and the Poisson brackets between the 
canonical variables:
\begin{eqnarray}
\label{eq:pbra3}
\{\pi^a,\phi^b\} &=& \delta^{ab}
\\
\label{eq:pbra4}
\{E_a^i,U_j\} & = & i t_a \delta^i_j U_j \nosum{j}
\\
\label{eq:pbra5}
\{E_a^i,E_b^j\} &=& \delta^{ij} f_{abc} E_c^j \nosum{j}.
\end{eqnarray}
The first one is the usual Poisson bracket between two canonically 
conjugate variables.
The second, \eq\nr{eq:pbra4} tells us that the transverse electric
field $E^i$ generates left translations on the SU(3) group manifold
of the link matrices $U_i$. It is actually more proper to consider
the Poisson bracket \eq\nr{eq:pbra4} as the definition of the 
electric field $E^i$. The relation between $E^i$ and the time 
derivative of the link matrix, \eq\nr{eq:Efield}
then follows  from this definition as the equation of motion for $U_i$.
The third Poisson bracket, \eq\nr{eq:pbra5}, 
follows directly from the second using the antisymmetry of the 
Poisson bracket and the Jacobi identity:
\begin{eqnarray}\label{eq:jacobi}
\{\{E_a^i,E_b^j\},U_k\} &=& 
-\{\{E_b^j,U_k\},E_a^i\} - \{\{U_k,E_a^i\},E_b^j\}
\\ \nonumber
&=&\delta^{ij} f_{abc}\{E_c^j,U_k\} \nosum{j}.
\end{eqnarray}

The equations of motion for the fields can be found by taking the
Poisson brackets between the fields and the Hamiltonian; the 
equation of motion for any quantity $v$ is $\dot{v} = \{\mathcal{H},v \}$. 
The equations of motion thus obtained are \cite{Lappi:2003bi,Krasnitz:1998ns}
\begin{eqnarray}\label{eq:mo1}
\dot{U}_i &=& i \frac{g^2}{\tau}E^i U_i \nosum{i=1,2}
\\
\label{eq:mo2}
\dot{\phi} &=& \tau \pi
\\
\label{eq:mo3}
\dot{E}^1 &=& \frac{i \tau}{2 g^2} \left[U_{1,2}+U_{1,-2} - \hc \right]
- \textrm{trace} + \frac{i}{\tau} [\tilde{\phi}_1,\phi] 
\\
\nonumber
\dot{E}^2 &=& \frac{i \tau}{2 g^2} \left[U_{2,1}+U_{2,-1} - \hc \right]
- \textrm{trace} + \frac{i}{\tau} [\tilde{\phi}_2,\phi] 
\\
\label{eq:mo4}
\dot{\pi} &=& \frac{1}{\tau}\sum_i\left[ 
\tilde{\phi}_i + \tilde{\phi}_{-i} - 2\phi \right],
\end{eqnarray}
where ``$- \textrm{trace}$'' means that the part proportional to the
identity matrix must be subtracted, because the electric fields $E^i$ 
are traceless matrices.
It is easy to check that these equations are invariant under gauge 
transformations that depend only on the transverse coordinates. Note that 
although the r.h.s. of the electric field equation of motion, \eq\nr{eq:mo3}
depends on different plaquettes, these are all based on the same point, which 
makes the equation gauge invariant.

\paragraph{Initial conditions on the lattice}
In addition to the equations of motion we also need the initial 
conditions corresponding to \eqs\nr{eq:trinitcond} and \nr{eq:longinitcond}
in the continuum. 
The initial condition for the link matrices $U_{(3)i}$ can be written as
\begin{equation}\label{eq:deltadelta2}
\tr \left[t_a \left[ 
\big(U^{(1)}_{i}+U^{(2)}_i \big)\big(1 + U^{\dag (3)}_{i}\big)
- \hc  \right] \right]=0,
\end{equation}
where $U_i^{(1,2)}$ are the pure gauge fields corresponding to the 
individual nuclei, given by
\begin{equation}
U_i^{(m)}(\xt)  = e^{i \Lambda_{(m)}(\xt)} e^{-i \Lambda_{(m)}(\xt+\itt)}
\textrm{, with }  \nabt^2 \Lambda_{(m)} = -g \rho_{(m)}.
\end{equation}
By expanding the link matrices in the limit $a \to 0$ it can be verified that
in the continuum limit this reduces to \eq\nr{eq:trinitcond}.
Equation~\nr{eq:deltadelta2} is a very nonlinear equation for $U_{(3)i}$, and must 
be solved with some kind of iterative process. When approaching the continuum 
limit the link matrices are closer to the identity matrix
and solving this equation numerically becomes simpler.
If  \eq\nr{eq:deltadelta2} is gauge transformed (with the gauge parameter
depending on the transverse coordinate), the identity matrix in 
$1 + U^{\dag(3)}_{i}$ becomes a pure gauge matrix. This can be 
interpreted\footnote{The interpretation is evident from the derivation in
\cite{Krasnitz:1998ns}.} so that the identity matrix corresponds to the gauge
field in the past light cone $x^+<0, \ x^-<0$. In this sense 
\eq\nr{eq:deltadelta2}
is \emph{not} gauge invariant, it applies in the specific, physically
well motivated, gauge where the field vanishes in the region of spacetime
where neither of the colliding nuclei has yet passed.

The initial condition for the longitudinal electric field can be written as
\begin{multline}\label{eq:latpiinit}
\pi(\xt) = \sum_{i} \frac{-1}{2g}\frac{i}{2}  \bigg[
\left(U_i^{(3)}(\xt) - 1\right)
\left(U_i^{\dag (2)}(\xt)-U_i^{\dag (1)}(\xt) \right) -\hc
\\ + 
\left(U_i^{\dag(3)}(\xt-\itt) - 1\right)
\left(U_i^{(2)}(\xt-\itt)-U_i^{(1)}(\xt-\itt) \right) -\hc
\bigg].
\end{multline}
It is easily verified that in the 
continuum limit \eq\nr{eq:latpiinit} reduces to the corresponding commutator 
equation in the continuum case, \eq\nr{eq:longinitcond}.
Once the more complicated initial condition for the transverse fields, 
\eq\nr{eq:deltadelta2}, has been solved, evaluating \eq\nr{eq:latpiinit} is
a straightforward calculation. 

As in the continuum case the initial conditions for the transverse electric 
fields and the  adjoint scalar field $\phi$ are
$E^i(\tau=0) = \phi(\tau=0) = 0$.

\paragraph{The leapfrog algorithm}

The leapfrog algorithm is commonly used in Hamiltonian time evolution of 
classical gauge theory. It is suited for partial differential equations
that are second order in time or, equivalently, for systems where the fields
(coordinates) and their time derivatives (momenta) are independent 
variables. The leapfrog algorithm is time reversal invariant, which guarantees 
second order accuracy in time.
The essential idea of this algorithm is that if the coordinates are
defined at times $\tau, \ \tau + \ud \tau, \ \tau+ 2\ud\tau, \dots$,
the momenta are defined at timesteps
$\tau + \half \ud \tau, \ \tau + \frac{3}{2}\ud \tau,
 \ \tau+ \frac{5}{2}\ud\tau, \dots$
The momentum at $\tau+ \half \ud \tau$ is then used to ``leap''
the coordinate from $\tau$ to $\tau+\ud \tau$, hence the name. 
The only peculiarity particular about the equations
of motion in this case, \eqs\nr{eq:mo1}--\nr{eq:mo4}, is the explicit time 
dependence, which must be treated properly in order to maintain the time
reversal symmetry of the algorithm. This simply means that when stepping
the fields from $\tau$ to $\tau + \ud \tau$ the explicit time argument
must be taken to be $\tau + \half \ud \tau$.

Explicitly, the timesteps can be written as
\begin{eqnarray}\label{eq:mo1d}
U_i(\tau+\ud \tau) &=& \exp\left( \frac{i g^2\ud \tau}{\tau + \ud \tau/2}
E^i(\tau+\ud \tau/2) \right) U_i(\tau) 
\\
\label{eq:mo2d}
\phi(\tau+\ud\tau) &=& \phi(\tau) +  (\tau+ \ud \tau/2) \ud \tau \pi(\tau + \ud \tau/2)
\\
\label{eq:mo3d}
E^1(\tau+\ud \tau) &=& E^1(\tau)
+ \frac{i (\tau+\ud \tau/2)\ud \tau}{2 g^2}
\left. \bigg[ 
U_{1,2}+U_{1,-2} - \hc - \textrm{trace} \bigg] \right|_{\tau+\ud \tau/2}
\nonumber \\  && 
+  \frac{i \ud \tau}{\tau+\ud \tau/2}
\left. \left[ \tilde{\phi}_1,\phi \right] \right|_{\tau+\ud \tau/2}
\\     
E^2(\tau+\ud \tau) &=& E^2(\tau)
+ \frac{i (\tau+\ud \tau/2)\ud \tau}{2 g^2}
\left. \bigg[ 
U_{2,1}+U_{2,-1} - \hc - \textrm{trace} \bigg] \right|_{\tau+\ud \tau/2}
\nonumber \\  \nonumber  && 
+  \frac{i \ud \tau}{\tau+\ud \tau/2}
\left. \left[\tilde{\phi}_2,\phi\right] \right|_{\tau+\ud \tau/2}
\\
\label{eq:mo4d}
\pi(\tau+\ud \tau) &=& \pi(\tau) + \frac{\ud \tau }{\tau+\ud \tau/2}\sum_i
\left. \left[ 
\tilde{\phi}_i + \tilde{\phi}_{-i}
 - 2\phi \right] \right|_{\tau+\ud \tau/2},
\end{eqnarray}
where the parallel transported scalar field $\tilde{\phi}_i$ was defined
in \eq\nr{eq:fitilde} and the index $i=1,2$ is not summed in \eq\nr{eq:mo1d}.

\section{Discretizing the Dirac equation in curved coordinates}
\label{sec:implicit}

\paragraph{Implicit discretization}

Let us first discuss the general difference between an implicit and an explicit 
discretization \cite{numrec}.
Consider a partial differential equation of first order in 
time\footnote{This is quite general, since e.g. a second order equation is
equivalent to a system of first order equations.}. Denote the values of the
unknown function at a timestep $n$ by $\xi_n,$ where $\xi_n$ is a vector
whose components are the values of the function at different points in space,
labeled by, say, $j$.
The partial differential equation can be written as
$\dot{\xi} = D \xi$, where $D$ is some 
differential operator, i.e. a matrix in position space $D_{jj'}.$
This equation can be discretized as $\xi_{n+1} = (1+\ud t D) \xi_n.$
This is an \emph{explicit} discretization, because only the spatial derivative
of the known function $\xi_n$ at the previous timestep is needed.
If might happen that the matrix $(1+\ud t D)_{jj'} = \delta_{jj'} + \ud t D_{jj'}$
has an eigenvalue with absolute value $> 1$. In this case the discretization is 
unstable, even if the original equation is not. This normally gives an upper limit, 
the Courant-Friedrichs-Lewy (CFL) condition, for the size of the timestep $\ud t$.
An example of an \emph{implicit} discretization, on the other hand, would
be $(1-\ud t D) \xi_{n+1} =  \xi_n,$ requiring one to solve a system of equations
(a linear one if $D$ is independent of the $\xi$) to find $\xi_{n+1}$ at each timestep.
The two discretization schemes are formally equally accurate (to first order
in $\ud t$), but the implicit one is usually stable for reasonable differential
operators $D.$

We shall now proceed to discretizing the Dirac equation in $\tau,z$-coordinates
implicitly and showing in detail how the resulting linear system can be solved
using LU-decomposition \cite{numrec}. Let us note, in passing, 
that it might well turn out
that to solve the 3+1-dimensional Yang-Mills equations in a coordinate system
where the time variable is $\tau,$ one will probably also have to use an implicit
discretization also. This is more difficult than the present case, because  unlike
the Dirac equation, the equations are not linear. Nevertheless, an implicit
scheme has already been used in Yang-Mills theory \cite{Tranberg:2003gi},
so the task should not be impossible.

\paragraph{Dirac equation in $\tau,z$-coordinates}

It is useful to separate the Dirac spinor $\psi$ into eigenvectors
of $\gt \gz$, by the projection operators $P ^\pm = \half (1\pm\gt\gz )$.
In the $\tau,z$-coordinate system and for these
 components separately the Dirac equation with an external gauge field is:
\begin{equation}\label{eq:diractauz}
\dtau \psi^\pm = \frac{\sqrt{\tau^2+z^2} \pm  z}{\tau}
\left( \mp \partial_z \psi^\pm + i \gt (i\gtrans \cdot \Dt - m) \psi^\mp \right)
\mp i  \frac{\phi}{\tau}\psi^\pm.
\end{equation}
Here we are using the same gauge as in the numerical calculation of gluon production,
namely $A_\tau=0$. We also assume that the background gauge field is independent of
rapidity and thus the longitudinal gauge field is an adjoint scalar field. 
From now on we absorb the 
coupling constant into the field and denote $\phi \equiv g A_\eta$. To shorten
the notation we shall also denote the timestep with a subscript $n$ and the 
lattice site in the longitudinal $z$-direction with a superscript $j$.
We discretize the Dirac equation implicitly as
\begin{multline}
\frac{1}{2 \ud \tau }\left(P^\pm \psi^j_{n+1}-P^\pm \psi^j_{n-1} \right)
\\
=
\mp 
\frac{\sqrt{\tau^2+z^2} \pm  z}{4 \ud z \tau}
\left[ P^\pm\psi^{j+1}_{n+1} +P^\pm\psi^{j+1}_{n-1}
 - P^\pm\psi^{j-1}_{n+1} -P^\pm\psi^{j-1}_{n-1} \right]
\\
+i \frac{\sqrt{\tau^2+z^2} \pm  z}{\tau} \gt (i\gtrans \cdot \Dt - m)
P^\mp \psi^j_n
\mp i  \frac{\phi_n}{2 \tau} P^\pm \left[ \psi^j_{n+1} + \psi^j_{n-1}  \right] 
\end{multline}
Multiplying by $2 \ud \tau$ and solving we get
\begin{multline}\label{eq:discrpsi}
\left( 1 \pm i  \frac{\phi_n \ud \tau}{ \tau}\right) P^\pm \psi^j_{n+1}
\pm \frac{\sqrt{\tau^2+z^2} \pm  z}{2 \ud z \tau} \ud \tau
\left[ P^\pm\psi^{j+1}_{n+1}  - P^\pm\psi^{j-1}_{n+1} \right]
\\
=
\left(1 \mp i  \frac{\phi_n \ud \tau}{ \tau}\right)
P^\pm \psi^j_{n-1}
\mp \frac{\sqrt{\tau^2+z^2} \pm  z}{2 \ud z \tau} \ud \tau
\left[ P^\pm\psi^{j+1}_{n-1}  - P^\pm\psi^{j-1}_{n-1} \right]
\\
- \frac{\sqrt{\tau^2+z^2} \pm  z}{\tau} 2 \ud \tau \gt (\gtrans \cdot \Dt + i m)
P^\mp \psi^j_n
\equiv \xi^j_n.
\end{multline}
Now we must address the question of the boundaries in the $z$-direction. Periodic
boundary conditions can be used in the transverse direction, but not for the
longitudinal coordinate, because the equation is not invariant under translations
in the $z$-direction at fixed \emph{proper} time $\tau$. We choose free boundary
conditions, not setting any restriction on $\psi$ at the boundary. In 
practice this means the following. Let us denote the endpoints of the lattice
by $j=\pm J$, i.e. the longitudinal lattice consists of $2J+1$ points. When for
the points in the interior of the lattice the longitudinal derivative was discretized
using the difference $\psi^{j+1}-\psi^{j-1}\approx 2 \ud z (\partial_z \psi)^j$, 
for the endpoints this must be replaced by 
$3 \psi^J +\psi^{J-2}-4\psi^{J-1} \approx 2 \ud z (\partial_z \psi)^J$
and $ 4 \psi^{-J+1}- 3 \psi^{-J} -\psi^{-J+2}\approx 2 \ud z (\partial_z \psi)^{-J}$.

Equation~\nr{eq:discrpsi} gives us a system of linear equations to solve at
each timestep. The coefficients of the system are $3\times3$ complex matrices
because of the gauge fields in the equation. Note that although we need
the spinor at both timesteps $n-1$ and $n$ to find $\psi_{n+1}$,
the $P^\pm$ components decouple in such a way that we only need to store
the values of $\psi$ at one timestep, i.e. after using $P^+\psi_{n-1}$ 
and $P^-\psi_n$
to find $P^+\psi_{n+1}$ we can then forget $P^+\psi_{n-1}$ and move to the next 
timestep to solve for $P^-\psi_{n+2}$. The linear system \eq\nr{eq:discrpsi} is 
most efficiently solved by LU-decomposition in a way that we will review next.

\paragraph{Solving the tridiagonal system by LU-decomposition}
Let us write the linear system, \eq\nr{eq:discrpsi}, as 
\begin{equation}
M P^\pm\psi_{n+1} =\xi_n,
\end{equation}
where $\xi_n$ as defined by the r.h.s. of \eq\nr{eq:discrpsi} is known and
$P^\pm\psi_{n+1}$ is unknown. The coefficients of the matrix $M$ can be read
off \eq\nr{eq:discrpsi} taking into account the special treatment of 
the boundary in the $z$-direction. The matrix $M$ is almost tridiagonal,
meaning that apart from the two exceptional elements
$m_{13}$ and $m_{NN-2}$ coming from the boundary
its elements other than $(i,i-1), \ (i,i) \textrm{ and } (i,i+1)$ are zero.
For notational simplicity we shall neglect the spinorial ($\gamma$-matrix)
indices; the following calculation separates easily for the two 
components of $P^\pm \psi$ ($\psi$ has four complex components, $P^\pm \psi$
has two that can be separated e.g. into the eigenvectors of $\gamma^1$.)
The LU-decomposition of $M$ is defined as follows
\begin{multline}
M = 
\left( \begin{array}{ccccccccc}
m_{11} & m_{12} & m_{13} & 0 & & & \cdots & & \\
m_{21} & m_{22} & m_{23} & 0 & & & & & \\
0 & m_{32} & m_{33} & m_{34} & 0 &  & &  & \\
\vdots &   & \cdots &  & \ddots  & & & & \\
& & & & & 0 & m_{N-1 N-2}& m_{N-1 N-1}& m_{N-1 N}\\
& &  & & & 0 & m_{N N-2}& m_{N N-1}& m_{N N} \\
\end{array} \right)
\\ = LU
=
\left( \begin{array}{cccccccc}
1 & 0 & & & \cdots &&& \\
l_1 & 1 & 0 &&&&& \\
0 & l_2 & 1 & && && \\
&&& & \ddots & &  & \\
&&&& 0 & l_N & l_{N-1} & 1
\end{array} \right)
\times \\
\left( \begin{array}{cccccccccc}
d_1 & u_2 & u_1 & 0 & & \cdots&  & \\
0 & d_2 & u_3 & 0 && &&\\
 & 0 & d_3 & u_4  &0 &&& \\
&&& & & \ddots & & \\
&&  &&& & d_{N-1} & u_N\\
&& & && & 0 & d_N
\end{array} \right),
\end{multline}
where we have defined $N \equiv 2 J +1$.
The algorithm for finding the elements of the lower and upper triangular matrices
$L$ and $U$ is the following
\begin{enumerate}
\item Set $d_1 = m_{11},$ $u_1 = m_{13},$ $u_2 = m_{12}.$
\item Set $l_1 = m_{21}/d_1$ and $u_3 = m_{23}-l_1 u_1$.
\item Set $u_{m+1} = m_{m m+1}$ for all $m \geq 3 $.
\item Set $d_m = m_{mm}-l_{m-1}u_m$ and $l_m = m_{m+1 m}/d_m$ for all $m \geq 2 $.
\item The last row is again exceptional: when we have $\dots l_{N-2},$ 
$\dots d_{N-1}$ we set $l_N = m_{N N-2}/d_{N-2},$ 
$l_{N-1}= (m_{NN-1}-l_N u_{N-1})/d_{N-1}$ and $d_N = m_{NN}-l_{N-1}u_N.$
\end{enumerate}
Note that this can be done handily in place (without additional memory, 
replacing the elements of $M$ by the elements of $L$ and $U$).
The expressions like $m_{21}/d_1$ actually mean $m_{21}(d_1)^{-1}$,
because $d_1$ is a complex $3\times3$-matrix.

Having LU-decomposed the coefficient matrix we can solve the equation 
by backsubstitution.
The equation we are solving is $LU \psi = \xi.$ Let us 
denote $U \psi = \chi; \ L \chi = \xi.$ We first find $\chi$ by noting that 
$\chi_1 = \xi_1,$ $\chi_{n+1} = \xi_{n+1} - l_n \chi_n,$ \dots, except for
the last exceptional element $\chi_N = \xi_N - l_N \chi_{N-2} -l_{N-1} \chi_{N-1}.$
This is the lower backsubstitution. Then follows the upper backsubstitution, starting
from the end and progressing backwards:
$\psi_N = d_N^{-1} \chi_N,$ then $\psi_n = d_n^{-1}(\chi_n - u_{n+1} \psi_{n+1})$
\dots, 
until the first one $\psi_1 = d_1^{-1} (\chi_1 - u_2 \psi_2 -u_1 \psi_3 )$.

Equation \nr{eq:discrpsi} has now been solved and we can proceed to the next
timestep. The number of operations in this solution is linearly proportional 
to the number of lattice points, as in an explicit scheme. The constant of
proportionality is, however, higher, so the algorithm is slower by a constant
factor.

\chapter{Results}
\label{chap:results}

Let us then summarize the most important findings of \refs
\cite{Lappi:2003bi,Lappi:2004sf,Gelis:2004jp}.

\section{Gluon production at central rapidities}
 
In \refc \cite{Lappi:2003bi} the $\kt$-distribution, the integrated
multiplicity and transverse energy of gluons produced in the
classical field model were calculated. The results corrected
an error in the normalization in 
\cite{Krasnitz:2002mn,Krasnitz:2001qu}\footnote{For an explanation of
this normalization issue see the erratum \cite{Krasnitz:2003jw}.}
and were found to agree with the experimental data within the limits
given by the analysis in \se\ref{sec:enmult}.
\refc \cite{Lappi:2003bi} also discussed the gauge dependence of the results and the relation to
the weak field result of \cite{Kovner:1995ts,Gyulassy:1997vt}
(see \se\ref{sec:pertsol}). The phase space density of gluons in this calculation
was found to be lower than one would expect, meaning that the classical field 
approximation is really applicable only to the very low momentum modes
with $\ptt \lesssim g^2 \mu$. Collisions of finite size nuclei with different 
impact parameters and using a modified probability distribution in stead
of the original McLerran-Venugopalan form \eq\nr{eq:rhorho} were also studied.

\begin{figure}[tbh]
\begin{center}
\noindent
\includegraphics[width=0.49\textwidth]{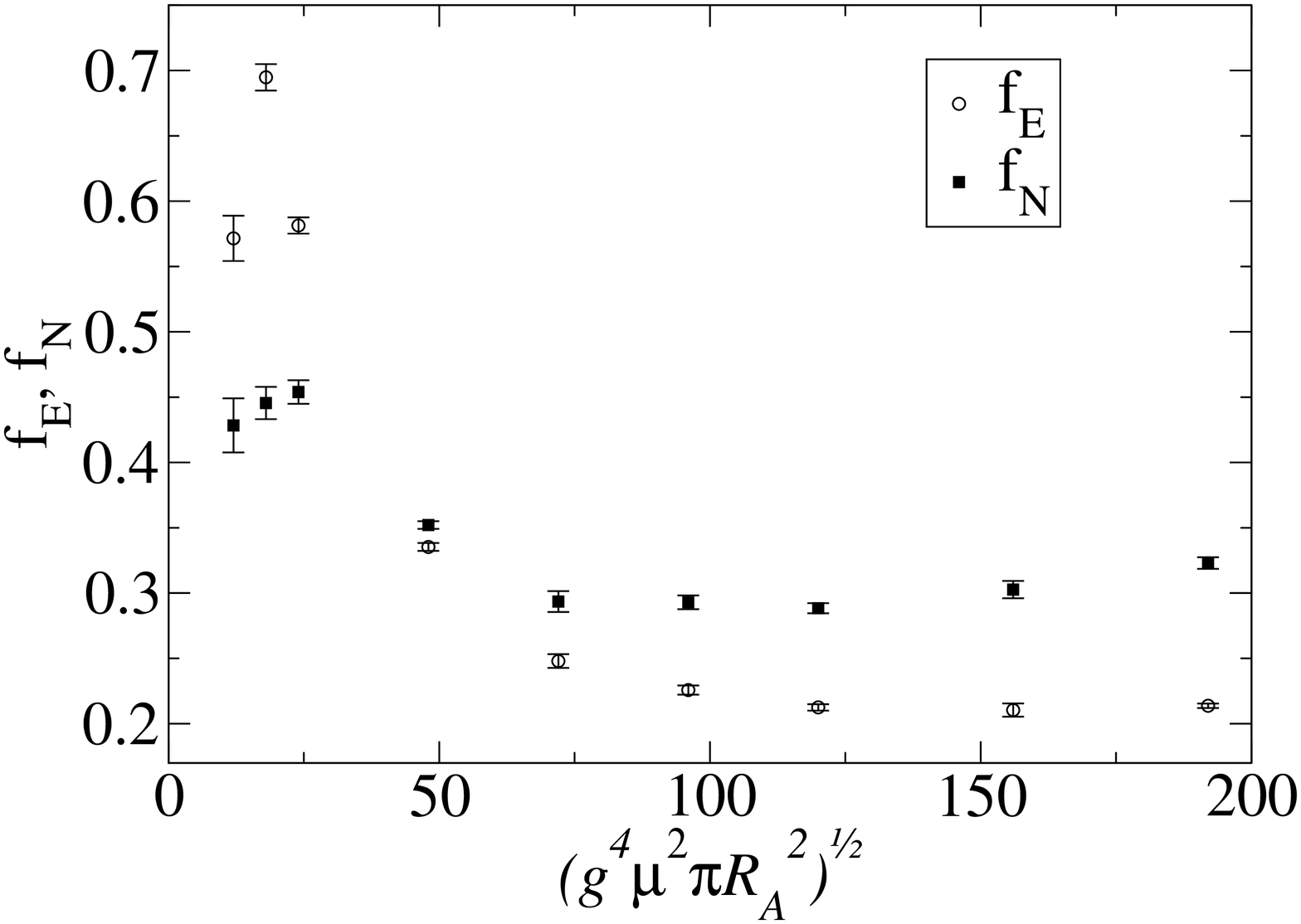}
\includegraphics[width=0.49\textwidth]{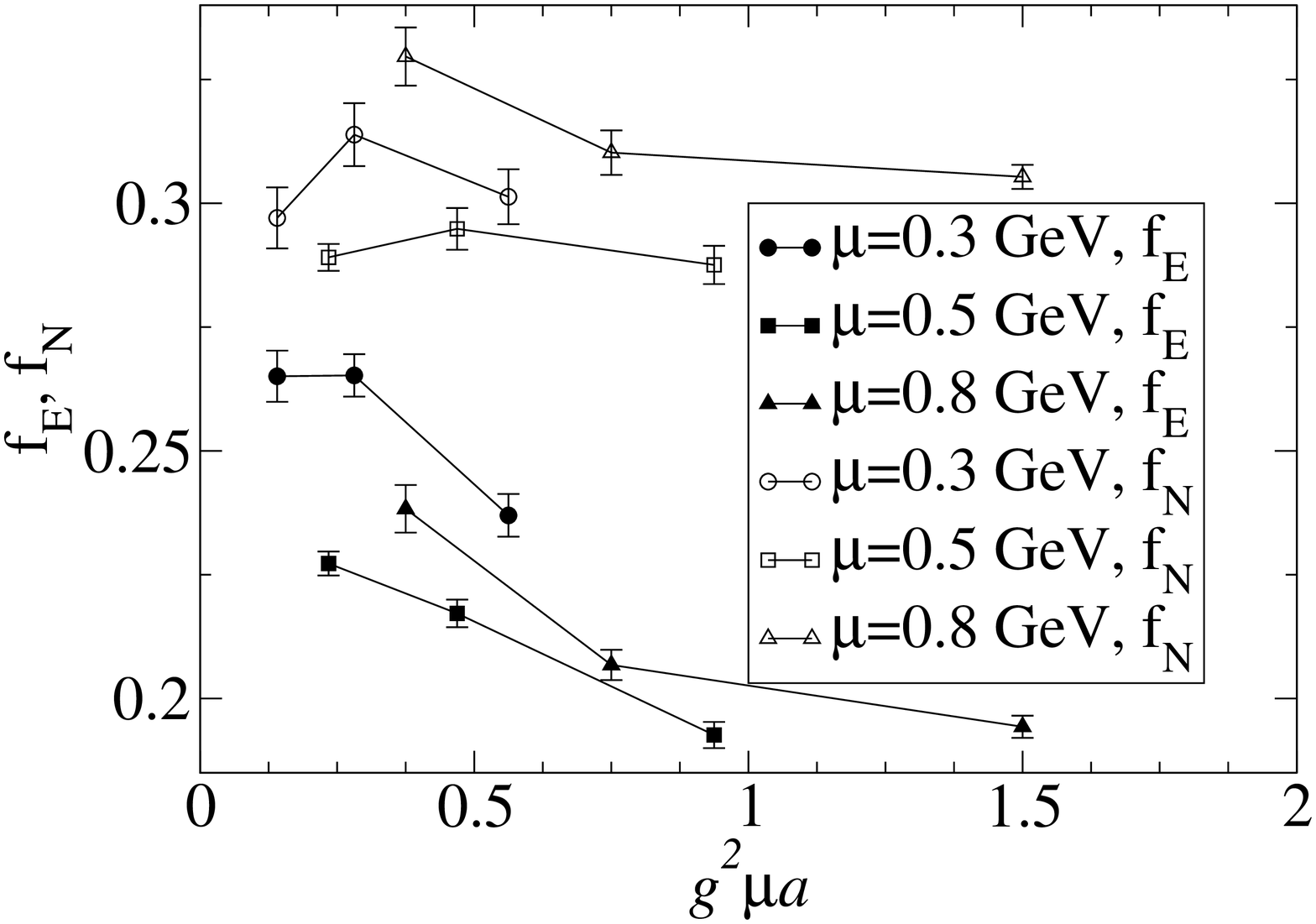}
\end{center}
\caption{Left: The dependence of $f_E$ and $f_N$ on the field strength
parameter $g^2 \mu \ra$. Right: The dependence of $f_E$ and $f_N$
on lattice spacing for different values of $\mu$.}\label{fig:saturcontlim}
\end{figure}

The energy and multiplicity, parametrized by the 
two dimensionless ratios
\begin{equation}
f_E = \frac{\ud E / \ud \eta}{g^4 \mu^3 \pi \ra^2 }
\quad\textrm{and}\quad
f_N = \frac{\ud N / \ud \eta}{g^2 \mu^2 \pi \ra^2 }
\end{equation}
were calculated numerically using the method outlined in \se\ref{sec:ccd}.
These ratios  were found to be approximately independent
of the parameters of the calculation ($g$, $\mu$, $\ra$ and the lattice 
spacing $a$) in the strong field limit ($g^2 \mu$ large enough),
as can be seen from \fig\ref{fig:saturcontlim}.

The values found in \refc \cite{Lappi:2003bi} were $f_E \approx 0.23$ and $f_N \approx 0.29$.
These fit in with the two phenomenological 
scenarios outlined in \se\ref{sec:enmult} in the 
following way. Assuming a nuclear transverse area of $140 \fm^2$, $g=2$ and
taking $g^2 \mu = 2.1 \gev$ one gets 
\begin{equation}
\frac{\ud N^\mathrm{init}_\mathrm{tot}}{\ud \eta} \approx
1100 \quad\textrm{and}\quad
\frac{\ud E^\mathrm{init}_T}{\ud \eta} \approx 1900 \gev,
\end{equation}
which fits in well with the hydrodynamical expansion scenario outlined in
\se\ref{sec:enmult}. On the other hand a choice of $g^2 \mu  = 1.4 \gev$
would give 
\begin{equation}\label{eq:hydromulti3}
\frac{\ud N^\mathrm{init}_\mathrm{tot}}{\ud \eta} \approx
500 \quad\textrm{and}\quad
\frac{\ud E^\mathrm{init}_T}{\ud \eta} \approx 550 \gev,
\end{equation}
which would be consistent with with the free streaming scenario, assuming
a factor of 2 increase in the multiplicity from the scatterings of the
partons and hadronization.

The CPU-time used for producing the results of \refc \cite{Lappi:2003bi} was of the order of
a few thousand hours on a 2.4 GHz Pentium processor, with comparable amounts
of computer resources used in the development phase of the code. The calculations 
were performed during the winter 2002-2003
using up to five 600 MHz Alpha EV6 processors in the ``dynamo'' 
minisupercluster at the Accelerator Laboratory
and three desktop 2.4 GHz Pentium machines
at the Theory Division, both at the University of Helsinki Department of Physics.
The program consisted of approximately 7000 lines of code in ANSI C in addition
to general purpose libraries for elementary complex number and SU(3)-matrix 
operations.

\section{Rapidity dependence}

\begin{figure}[tbh]
\begin{center}
\noindent
\includegraphics[width=0.7\textwidth]{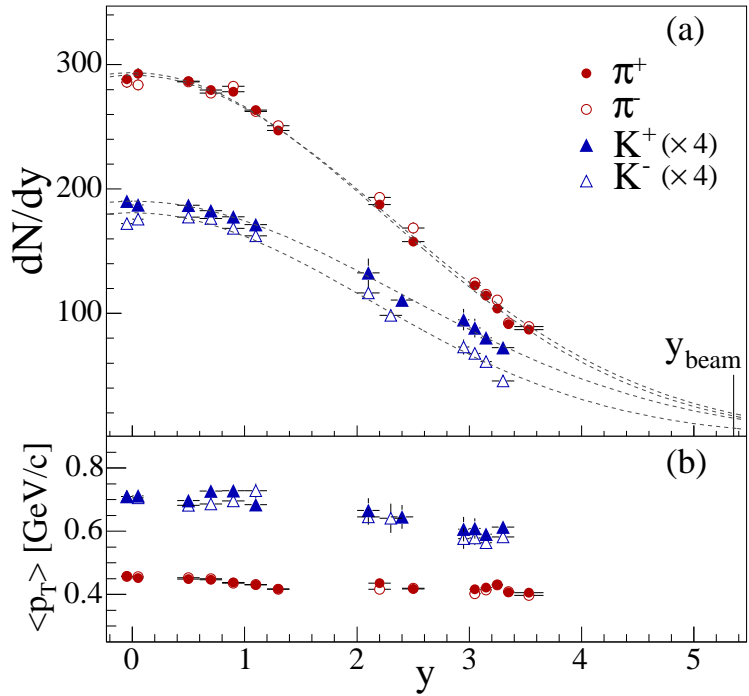}
\end{center}
\caption{BRAHMS results on charged meason rapidity distributions in 
5~\% most central
gold-gold collisions \protect\cite{Bearden:2004yx}. The dashed lines are Gaussian
fits to the data with $\sigma_{\pi^+} = 2.25 \pm 0.02$ (stat)
and $\sigma_{\pi^-} = 2.29 \pm 0.02$ (stat). The lower panel
shows the mean transverse momentum of the mesons for different rapidities.
}\label{fig:brahms}
\end{figure}

The calculations of \refc \cite{Lappi:2003bi} were further extended in 
\refc \cite{Lappi:2004sf}. The experimentally found
rapidity (or $\ln 1/x$) dependence of the saturation scale discussed
in \se\ref{sec:dis} was exploited in a simple extension of the 
calculation of \cite{Lappi:2003bi} to study the rapidity 
dependence of gluon production. The main assumption of the calculation
was the following: in the fully boost invariant model described in 
\se\ref{sec:kmclw} the only (implicit) dependence on rapidity comes
though the strengths of the classical color sources $\rho_{(1,2)}$.
It has been experimentally observed that the saturation scale
varies with rapidity as $\qs^2(x) \sim \xbj^{-\lambda}$. The saturation
scale in the McLerran-Venugopalan model is proportional, up to a logarithm,
to the source strength, $\qs \sim g^2 \mu (\times \ln g^2\mu).$
It was argued in \refc \cite{Lappi:2004sf} that for central rapidities in heavy ion 
collisions the dominant $\xbj$ scale that should be used to evaluate
the saturation scale is given by $\xbj_{1,2} \sim e^{\pm y}$ for the
two nuclei. One can then study gluon production at the rapidity $y$ 
by choosing the source strengths as $g^4 \mu^2 \sim e^{\pm \lambda y}.$

The result of \refc \cite{Lappi:2004sf} was that the multiplicity
of produced gluons can be described by very broad Gaussians in $y$ with width
$\sigma \approx 6$. This kind of a Gaussian is much broader than the one
observed at RHIC energies by the BRAHMS collaboration
 \cite{Bearden:2004yx}\footnote{The 
rapidity dependence has also been measured by STAR \cite{Adams:2003xp}, but
the coverage in rapidity is too small ($|y| < 0.5$) to give a clear
interpretation.} (see \fig\ref{fig:brahms}),
 but coincidentally close to the pQCD+saturation
model result \cite{Eskola:2002qz,Tuominen:2002sq}\footnote{Note that according
to \refc\cite{Eskola:1997hz} hydrodynamical evolution does 
not change the rapidity distribution enough to explain this large a difference
between the initial and final states.}. The conclusion
suggested in \refc \cite{Lappi:2004sf} was then that at RHIC energies the rapidity
dependence of particle production at central rapidities at AuAu-collisions
at RHIC energies is not dominated by saturation physics. It is perhaps
more dependent on the kind of high $\xbj$ physics (the $(1-x)^4$-behavior
of the gluon structure functions expected from momentum sum rules) incorporated
e.g. in the ``saturation model'' calculations of 
\refs\cite{Kharzeev:2001gp,Hirano:2004rsb,Hirano:2004rs} to
reproduce the experimentally observed rapidity (or pseudorapidity)
dependence.

The CPU-time used for producing the results of \refc \cite{Lappi:2004sf} was larger than
what was used for  \refc \cite{Lappi:2003bi} because the computations had to be done separately
for different values of the rapidity. The total CPU-time was 
approximately 5000 hours on a 2.4 GHz Pentium processor. The calculations were
performed during the late summer of 2004
mostly using three desktop 2.4 GHz Pentium machines
at the Theory Division of the University of Helsinki Department of Physics.

\section{Quark pair production: 1+1-dimensional toy model}

In \refc \cite{Gelis:2004jp} the calculation of
pair production by classical fields was formulated based
on the theory detailed in \refc\cite{Baltz:2001dp} and closely following the
calculation in the Abelian theory performed in \refc\cite{Baltz:1998zb}.
It was then argued that the practical numerical solution of the Dirac equation
in the forward light cone with the initial condition given at $\tau=0$
is most practically performed using as coodinates $\tau$ and $z$.
The implicit discretization scheme required for this numerical solution
was then constructed and demonstrated in a 1+1-dimensional toy model.

\begin{figure}[!t]
\begin{center}
\includegraphics[width= 0.7\textwidth]{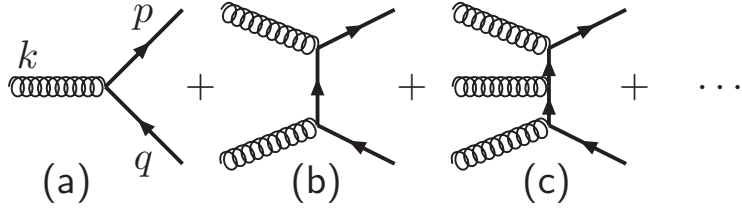}
\end{center}
\caption{Diagrams contributing to the quark pair prodction amplitude in the
1+1-dimensional toy model.}\label{fig:diags}
\end{figure}

\begin{figure}[!b]
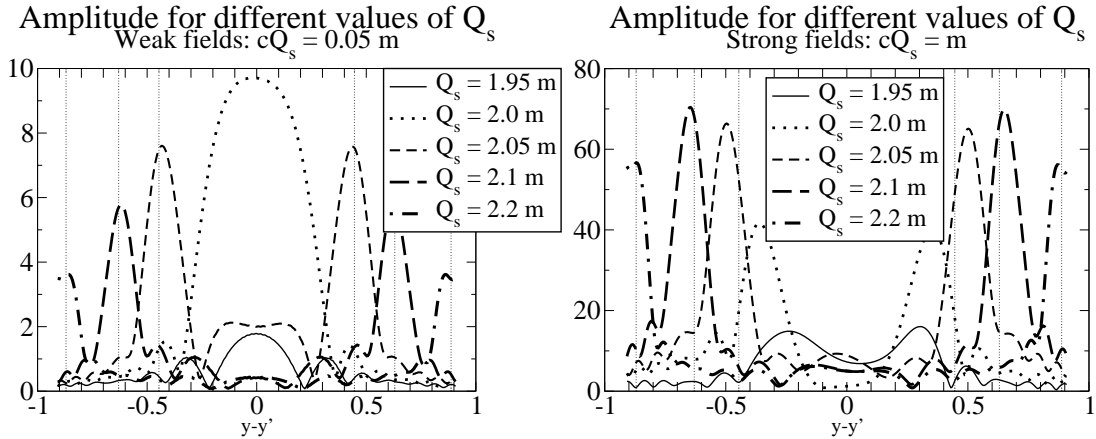

\begin{center}
\noindent
\includegraphics[width=0.50583\textwidth]{pertpeak.eps}
\includegraphics[width=0.47417\textwidth]{strongfieldpeak.eps}
\end{center}
\caption{Absolute value of the quark pair production
amplitude for different values of the oscillation scale $\qs$. Left:
weak fields, the peaks are at the location given by the lowest order 
perturbative result.
Right: strong fields with the same values of $\qs.$ Peaks near
``threshold'' $\qs=2m$ are shifted.  } \label{fig:ampli}
\end{figure}

In the gauge $A_\tau=0$ that was used in calculating the classical 
background field the Dirac equation of the 1+1-dimensional toy model is
\begin{equation}\label{eq:eomtauz}
\dtau \psi = \frac{\sqrt{\tau^2+z^2}+ \gt \gz z}{\tau}
\left(-\gt \gz \partial_z \psi 
- im_\textrm{eff} \gt \psi \right) - i \gt \gz   g \frac{A_\eta}{\tau} \psi.
\end{equation}
The transverse coordinate dependence has been neglected and the 
mass of the 1+1-dimensional theory corresponds to the transverse mass of the
full theory, $m^2_\textrm{eff} \approx \kt^2 + m^2$. The 1+1-dimensional 
toy model in the $A_\tau=0$-gauge only contains one component of the gauge 
field, $A_\eta$. In  \refc \cite{Gelis:2004jp} this model was studied using two forms for the time 
dependence of this background field, an exponential decay and a Bessel
function:
\begin{equation}\label{eq:besseleta}
A_\eta = c \qs \tau J_1 (\qs \tau).
\end{equation}
This is the correct time dependence of the 
perturbative solution to the gauge field equations of motion \cite{Kovner:1995ts}.

For weak fields $c \ll 1$ the amplitude for quark pair production 
can be computed
using the first order in perturbation theory (diagram (a) in Fig.~\ref{fig:diags}).
The result for the oscillating field of \eq\nr{eq:besseleta} is a peak at
\begin{equation}\label{eq:resonans}
2k^+ k^- = (p+q)^2 = 2m_\textrm{eff}^2 \left(1 + \cosh (\Delta y) \right) =  \qs^2.
\end{equation}
As is demonstrated in  \fig\ref{fig:ampli}, for weak fields the
 numerical computation of \refc \cite{Gelis:2004jp}
reproduces this perturbative peak. For stronger fields the numerical 
solution sums over all the diagrams in \fig\ref{fig:diags} and  the position 
of the peak is shifted.
The appearence of this peak is not a physical effect. The time dependent
field of \eq\nr{eq:besseleta} represents off-shell gluons with invariant mass
$\qs^2$ decaying into quark pairs, and the amplitude has a peak where
the invariant mass of the pair equals that of the gluon. 
In the full 3+1-dimensional case the gluons are on shell, and
this peak is washed out by integration
over the relative transverse momentum of the pair.
The effect serves here as a verification that the numerical
method for solving the Dirac
equation and projecting out the positive energy solutions works.

\chapter{Review and outlook}
\label{chap:conclusion}

The classical field approximation is a theoretically appealing way
to study QCD as it manifests itself in the small $\xbj$ proton or nucleus 
wavefunction and in relativistic heavy ion collisions. The aim of this thesis
has been to review how a classical field model can be used to understand 
the initial stages of a heavy ion collision towards thermalization, 
starting from a model of the nuclear wavefunction. It has been seen
that the classical field model gives a plausible picture of gluon production
in the central rapidity region of a heavy ion collision. One can numerically
calculate the gluon multiplicity and transverse energy created in the initial
stages of the collision. This calculation was the subject of the 
first article of this thesis \cite{Lappi:2003bi}, and it was shown that
the values obtained are compatible with 
experimental observations within the Bjorken scenario of boost invariant 
ideal hydrodynamical expansion.

Due to the uncertainty of the appropriate
value of the source charge density the results of the classical model do
not, however, rule out a free-streaming scenario without the decrease of 
energy density due to  $p\ud V$-work in the hydronynamical evolution. 
This uncertainty can be reduced in two ways. Firstly observations of 
the hadron suppression at large $\ptt$, the disappearence of away-side 
jet-like correlations and elliptic flow favor the interpretation in terms
of the Bjorken scenario, although hydrodynamical models do have difficulties
with HBT radii and there have been attempts to explain elliptic flow without
referring to hydrodynamics \cite{Kovchegov:2002nf,Krasnitz:2002ng}. 
 Secondly one should in principle be able to
independently determine the value of the source strength parameter using
data from deep inelastic scattering. Doing this accurately enough 
to clearly distinguish between $g^2 \mu = 1.4 \gev$ and $g^2 \mu = 2.1 \gev$
is, however, not easy.

Inclusion of  the longitudinal dimension is important for fully understanding
two seemingly unrelated issues. One is the thermalization, or specifically
the isotropisation, of the initial gluonic system. The essential assumption
of the hydrodynamical model is that the thermal system is isotropic, i.e.
has not only transverse but also longitudinal pressure. Explaining the creation of 
longitudinal pressure, if it can be done in the classical field model, must 
necessarily involve a full 3+1-dimensional solution of the gauge field equations 
of motion. The second issue that requires a better treatment of the longitudinal
dimension is including the effects of JIMWLK evolution. The approach used
in the second article of the present thesis
\cite{Lappi:2004sf}, consists of varying the initial conditions of the 
same 2+1-dimensional calculation to account for the change in the saturation
scale at different rapidities. There exists a large
body of both analytical and numerical work on this equation, which is not yet
fully included in this simple approach.

The question of theoretically understanding of the thermal and chemical 
equilibration that the experimental results seem to indicate is not as
well understood. In the third article of this thesis
we have formulated  a way to calculate the
number of quark pairs produced in the classical field model. The calculation 
essentially consists of solving the Dirac equation in the background
given by the classical gauge fields.  Many technical  issues concerning,
in particular, the longitudinal dimension have been solved in a 1+1-dimensional
toy model. The full 3+1-dimensional case, requiring an extensive 
numerical calculation, has been formulated but not yet implemented numerically.

\subsection*{Acknowledgements}

I want to thank K. Kajantie for directing me to this direction of research and 
advice, K. J. Eskola, F. Gelis and K. Rummukainen for reading this manuscript,
and P. Hoyer and P.V. Ruuskanen for discussions.
This work has been supported by the Finnish Cultural Foundation
and the Magnus Ehrnrooth Foundation as well as
the Helsinki Institute of Physics and the Academy of Finland,
contract no. 77744.

\appendix

\chapter{Notations}

\section{Spacetime} \label{sec:spacetime}

The metric in the  original $\{t, z,\xt \}$-basis is 
$g_{\mu \nu} = \textrm{diag}(1, -1, -1,-1).$ Thus
$\partial_i \partial^i  = -  \partial_i \partial_i = - \nabt^2$.
We define the light cone coordinates $x^\pm$ and proper time and 
spacetime rapidity $\tau,\eta$ as:
\begin{equation}
\begin{aligned}
x^{\pm} && = && \frac{1}{\sqrt{2}}(t \pm z) && = && 
\frac{1}{\sqrt{2}} \tau e^{\pm \eta}
\\
\tau &&=&& \sqrt{t^2-z^2} &&=&& \sqrt{2x^{+} x^{-}} 
\\
\eta &&=&& \frac{1}{2}\ln{\frac{t+z}{t-z}}
&&=&& \frac{1}{2}\ln{\frac{x^{+}}{x^{-}}},
\end{aligned}
\end{equation}
which gives the original coordinates in terms of the new ones as
\begin{equation}
\begin{aligned}
t && = && \frac{1}{\sqrt{2}}(x^+ + x^-) && = && \tau \cosh{\eta}
\\
z && = && \frac{1}{\sqrt{2}}(x^+ - x^-) && = && \tau \sinh{\eta}. 
\end{aligned}
\end{equation}
The metric in the light cone coordinates is
$\ud s^2 = 2 \ud x^+ x^- - \ud \xt^2 $
and the $\tau, \eta$ coordinates 
$\ud s^2 = \ud \tau^2 - \tau^2 \ud \eta^2 - \ud \xt^2,$
giving the invariant integration measure
$\sqrt{|g|} \ud^4 x = \ud z \ud t \ud^2 \xt 
= \ud x^+ \ud x^-  \ud^2 \xt
= \tau \ud \eta \ud \tau \ud^2 \xt.$
Any vector, in particular the gauge field $A_\mu$, transforms
under coordinate transformations in the familiar way
\begin{equation}\label{eq:vectransf}
A^{\mu'} = \frac{\partial x^{\mu'}}{\partial x^\mu}  A^\mu \quad
A_{\mu'} = \frac{\partial x^\mu}{\partial x^{\mu'}}  A_\mu.
\end{equation}

Straightforward application of \eq\nr{eq:vectransf} gives the 
$\tau,\eta$-components of a vector as
\begin{equation}
\begin{aligned}
A_\tau &&=&& A^\tau &&=&&(x^+A^- + x^-A^+)/\tau \\
A_\eta &&=&&-\tau^2 A^\eta &&=&& x^+A^- - x^-A^+.
\end{aligned}
\end{equation}

\section{Chromodynamics}\label{sec:chromo}

We denote the generators of the fundamental representation of the gauge group
by $t_{a}$. They are related to the Gell-Mann matrices $\lambda^a$ and
the Pauli matrices $\sigma^a$ by $t_a = \lambda^a/2$ for $\nc=3$ and  
$t_a=\sigma_a/2$ for $\nc=2$. 
The generators are normalized as
\begin{equation}\label{eq:normalisaatio}
\tr t_a t_b = \frac{1}{2} \delta_{ab}
\end{equation} 
\begin{equation}
[t_a,t_b] = i f_{abc}t_c.
\end{equation} 
The structure constants 
are completely antisymmetric in their indices \cite{weinberg2}.
The color indices $a,b,c, \dots$ range from 1 to $\nc^2-1$, the dimension 
of the group. They are written here indifferently as superscripts or 
subscripts.
The adjoint representation is generated by $(T_a)_{bc} = -if_{abc}$.
\begin{equation}\label{eq:adjnormalisaatio}
\tr T_a T_b = \nc \delta_{ab}, 
\textrm{ or } T_a T_a = \nc \mathbf{1}_{(\nc^2-1)},
\end{equation} 
where $\mathbf{1}_{(\nc^2-1)}$ is the $(\nc^2-1)\times(\nc^2-1)$ identity 
matrix.
The adjoint representation of an SU($\nc$)-matrix $U=\exp{i\chi^a t_a}$ is
denoted by $\dadj(U) \equiv \exp{i\chi^a T_a}$, and is related to the
fundamental representation by
\begin{equation}\label{eq:adjes2}
2 \tr ( t_a U^\dag t_b U ) = (\dadj(U))_{ab}.
\end{equation}
We use the following notation for the dimensions and the Casimirs of the 
representations:
\begin{equation}
\begin{array}{rclcrcl}
\df &=& \nc & \  & \da & = &\nc^2 - 1
\\ 
\cf &=& \frac{\nc^2-1}{2\nc} & \ & \ca & =&\nc\\
\end{array}
\end{equation}

The action of QCD is 
\begin{equation}\label{eq:qcdaction}
S = 
\intd \ud^4 x 
\sum_f\bar{\psi}_f (i D \!\!\!\! / \! - m  )\psi_f,
-\frac{1}{2} \tr F^{\mu \nu} F_{\mu \nu} 
\end{equation} 
with the covariant derivative
\begin{equation}\label{eq:covder}
D^{\mu} = \partial^{\mu} + i g A^{\mu},
\end{equation} 
giving the fields trength tensor
\begin{equation}\label{eq:fstrength}
F^{\mu \nu} = -\frac{i}{g}[D^\mu , D^\nu] =  
\partial^{\mu} A^{\nu} - \partial^{\nu} A^{\mu} + ig[A^{\mu},A^{\nu}]
\end{equation} 
or in components
\begin{equation}
F^{\mu \nu}_a =
\partial^{\mu} A^{\nu}_a - \partial^{\nu} A^{\mu}_a - g 
f^{abc}A^{\mu}_b A^{\nu}_c.
\end{equation} 

QCD is invariant under gauge transformations, parametrized by 
a spacetime dependent matrix in the gauge group, $U(x)$:
\begin{equation}\label{eq:gaugetransf}
A^{\mu} \to UA^{\mu}U^\dag  -\frac{i}{g}U\partial^{\mu}U^\dag.
\end{equation}
Under these transformations the covariant derivative and the 
field strength tensor transform as
\begin{equation}\label{eq:gaugetransf2}
D^{\mu} \to U D^{\mu} U^\dag \ \textrm{ and } \
F^{\mu \nu} \to U F^{\mu \nu} U^\dag.
\end{equation}
A field configuration that is gauge equivalent to the vacuum is called 
a pure gauge:
\begin{equation}\label{eq:puregauge}
A^{\mu} = -\frac{i}{g}U\partial^{\mu}U^\dag.
\end{equation}

\section{Gauge fields on the lattice}
\label{sec:wilson}

\begin{figure}[!h]
\begin{center}
\noindent
\includegraphics{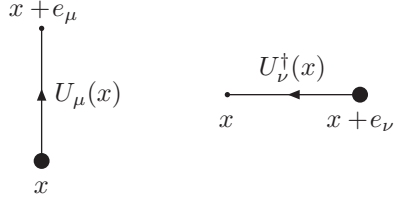}
\end{center}
\caption{ 
The link matrix $U_\mu(x)$ connects the lattice site $x$ to $x+\emu$ (left). 
The Hermitian conjugate $U^\dag_\nu(x)$ connects lattice points in the 
negative direction, going from $x+ \enu$ to $x$.
}\label{fig:links}
\end{figure}

The fundamental object of lattice gauge field theory is the
link matrix $U_\mu(x)$ connecting the point $x$ to the neighboring
lattice site $x+e_\mu$ (see \fig\ref{fig:links}).
The Hermitian conjugate of the link matrix
reverses the direction, connecting $x+e_\mu$ to $x.$ In terms of
the continuum gauge field $A_\mu$ we identify the link matrix as
$U_\mu(x) = e^{igaA_\mu(x)}$ 

A gauge transformation on the lattice transforms the link matrices as
\begin{equation}\label{eq:gaugetrans}
U_\mu(x) \to V(x) U_\mu(x) V^\dag (x\!+\emu),
\end{equation}
and the fermion fields as
\begin{equation}
\psi(x) \to V(x) \psi(x), \quad \bar{\psi}(x) \to  \bar{\psi}(x)V^\dag, 
\end{equation}
leaving the following combination gauge-invariant
\begin{equation}\label{eq:gaugetrans2}
\bar{\psi}(x)\gamma^{\mu}U_\mu(x)\psi(x\!+\emu).
\end{equation}
A pure gauge field is one that can be gauge transformed to zero, meaning
that the link matrices can be transformed to the indentity matrix:
\begin{equation}\label{eq:hilapuregauge}
U_\mu(x) = V(x)V^\dag (x\!+\emu).
\end{equation}
 
 \begin{figure}[!h]
\begin{center}
\noindent
\includegraphics{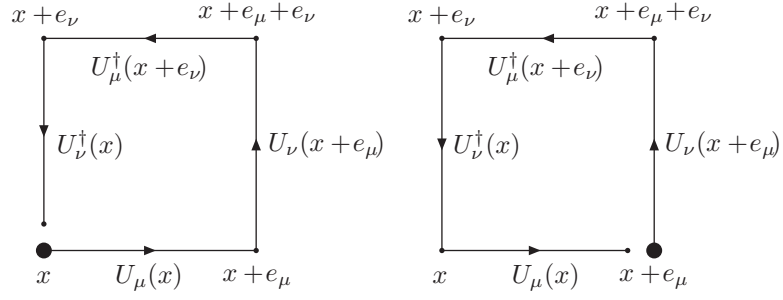}
\end{center}
\caption{ 
The plaquettes $U_{\mu,\nu}(x)$ (left) and $U_{\nu,-\mu}(x+\emu)$ (right)
circle the same contour, but gauge transform on different points, 
$x$ and $x+\emu$ respectively.
}\label{fig:plaqs}
\end{figure}

A gauge invariant action can be defined using the plaquette
\begin{equation}\label{eq:plaketti1}
U_{\mu, \nu}(x) =
 U_\mu(x) U_\nu(x\! + \emu) U^\dag _\mu(x\!+\enu) U^\dag _\nu(x).
\end{equation}
With a subscript $-\mu$ we denote the negative $\mu$-direction. Thus
a plaquette around the same contour, but based on a different point 
(see \fig\ref{fig:plaqs}) can be written e.g. as 
\begin{equation}
U_{\nu,-\mu}(x+\emu) = 
U_\nu(x\! + \emu) U^\dag _\mu(x\!+\enu) U^\dag _\nu(x) U_\mu(x).
\end{equation}
As can be seen from its definition and the transformation properties of
the link matrices, the plaquette gauge transforms at its base point
\begin{equation}
U_{\mu, \nu}(x)  \to V(x) U_{\mu, \nu}(x) V^\dag (x).
\end{equation}
Because of cyclicity, the \emph{trace} of the plaquette is 
gauge invariant and independent of the base point
\begin{equation}\label{eq:plaketti3}
\tr U_{\mu,\nu}(x)= \tr U_{\nu,-\mu}(x+\emu)= \tr U_{-\mu,-\nu}(x+\emu+\enu)
= \tr U_{-\nu,\mu}(x+\enu).
\end{equation}
In the continuum limit the plaquette becomes
\begin{equation}
U_{\mu,\nu} = e^{iga^2 F_{\mu\nu} + \mathcal{O}(a^3)}.
\end{equation}
and the pure gauge action $S = - \half \tr F_{\mu\nu}^2$ can be represented
by the \emph{Wilson action} \cite{Wilson:1974sk}
\begin{equation} \label{eq:mwapp}
S = - \frac{2\nc}{g^2}
\sum_x \sum_{\mu<\nu} \left( 1-\frac{1}{\nc}\R \ \tr U_{\mu,\nu} \right).
\end{equation}
In Minkowski space one must choose a different sign for the $0,i$-terms 
to reproduce the right metric.

\bibliographystyle{h-physrev4}
\bibliography{spires}

\begin{thebibliography}{100}

\bibitem{Lappi:2003bi}
T.~Lappi,
\newblock Phys. Rev. {\bf C67}, 054903 (2003), [hep-ph/0303076].

\bibitem{Lappi:2004sf}
T.~Lappi,
\newblock Phys. Rev. {\bf C70}, 054905 (2004), [hep-ph/0409328].

\bibitem{Gelis:2004jp}
F.~Gelis, K.~Kajantie and T.~Lappi,
\newblock Phys. Rev. {\bf C71}, 024904 (2005), [hep-ph/0409058].

\bibitem{Adams:2005dq}
STAR, J.~Adams {\em et~al.},
\newblock nucl-ex/0501009.

\bibitem{Adcox:2004mh}
PHENIX, K.~Adcox {\em et~al.},
\newblock nucl-ex/0410003.

\bibitem{Arsene:2004fa}
BRAHMS, I.~Arsene {\em et~al.},
\newblock nucl-ex/0410020.

\bibitem{Back:2004je}
PHOBOS, B.~B. Back {\em et~al.},
\newblock nucl-ex/0410022.

\bibitem{Freund:2002ux}
A.~Freund, K.~Rummukainen, H.~Weigert and A.~Schafer,
\newblock Phys. Rev. Lett. {\bf 90}, 222002 (2003), [hep-ph/0210139].

\bibitem{Adcox:2001mf}
PHENIX, K.~Adcox {\em et~al.},
\newblock Phys. Rev. Lett. {\bf 88}, 242301 (2002), [nucl-ex/0112006].

\bibitem{Adler:2003kg}
PHENIX, S.~S. Adler {\em et~al.},
\newblock Phys. Rev. Lett. {\bf 91}, 172301 (2003), [nucl-ex/0305036].

\bibitem{Adler:2003cb}
PHENIX, S.~S. Adler {\em et~al.},
\newblock Phys. Rev. {\bf C69}, 034909 (2004), [nucl-ex/0307022].

\bibitem{Fries:2004ej}
R.~J. Fries,
\newblock J. Phys. {\bf G30}, S853 (2004), [nucl-th/0403036].

\bibitem{Ehtamo:1983hu}
H.~Ehtamo, J.~Lindfors and L.~D. McLerran,
\newblock Z. Phys. {\bf C18}, 341 (1983).

\bibitem{McLerran:1994ni}
L.~D. McLerran and R.~Venugopalan,
\newblock Phys. Rev. {\bf D49}, 2233 (1994), [hep-ph/9309289].

\bibitem{McLerran:1994ka}
L.~D. McLerran and R.~Venugopalan,
\newblock Phys. Rev. {\bf D49}, 3352 (1994), [hep-ph/9311205].

\bibitem{McLerran:1994vd}
L.~D. McLerran and R.~Venugopalan,
\newblock Phys. Rev. {\bf D50}, 2225 (1994), [hep-ph/9402335].

\bibitem{Kharzeev:2001gp}
D.~Kharzeev and E.~Levin,
\newblock Phys. Lett. {\bf B523}, 79 (2001), [nucl-th/0108006].

\bibitem{Kharzeev:2002pc}
D.~Kharzeev, E.~Levin and L.~McLerran,
\newblock Phys. Lett. {\bf B561}, 93 (2003), [hep-ph/0210332].

\bibitem{Kharzeev:2003wz}
D.~Kharzeev, Y.~V. Kovchegov and K.~Tuchin,
\newblock Phys. Rev. {\bf D68}, 094013 (2003), [hep-ph/0307037].

\bibitem{Kharzeev:2004if}
D.~Kharzeev, E.~Levin and M.~Nardi,
\newblock hep-ph/0408050.

\bibitem{Kharzeev:2002ei}
D.~Kharzeev, E.~Levin and M.~Nardi,
\newblock Nucl. Phys. {\bf A730}, 448 (2004), [hep-ph/0212316].

\bibitem{Blaizot:2004wu}
J.~P. Blaizot, F.~Gelis and R.~Venugopalan,
\newblock hep-ph/0402256.

\bibitem{Blaizot:2004wv}
J.~P. Blaizot, F.~Gelis and R.~Venugopalan,
\newblock hep-ph/0402257.

\bibitem{Kharzeev:2003sk}
D.~Kharzeev and K.~Tuchin,
\newblock Nucl. Phys. {\bf A735}, 248 (2004), [hep-ph/0310358].

\bibitem{Krasnitz:1998ns}
A.~Krasnitz and R.~Venugopalan,
\newblock Nucl. Phys. {\bf B557}, 237 (1999), [hep-ph/9809433].

\bibitem{Bass:1999zq}
S.~A. Bass {\em et~al.},
\newblock Nucl. Phys. {\bf A661}, 205 (1999), [nucl-th/9907090].

\bibitem{Lappi:2004xp}
T.~Lappi,
\newblock hep-ph/0409087.

\bibitem{Gyulassy:2004zy}
M.~Gyulassy and L.~McLerran,
\newblock nucl-th/0405013.

\bibitem{Blaizot:2004px}
J.-P. Blaizot and F.~Gelis,
\newblock hep-ph/0405305.

\bibitem{Bjorken:1982qr}
J.~D. Bjorken,
\newblock Phys. Rev. {\bf D27}, 140 (1983).

\bibitem{Blaizot:1987nc}
J.~P. Blaizot and A.~H. Mueller,
\newblock Nucl. Phys. {\bf B289}, 847 (1987).

\bibitem{Blaizot:2003tw}
J.-P. Blaizot, E.~Iancu and A.~Rebhan,
\newblock hep-ph/0303185.

\bibitem{Nagle:2002wj}
J.~L. Nagle and T.~S. Ullrich,
\newblock nucl-ex/0203007.

\bibitem{Schukraft:2001sv}
J.~Schukraft,
\newblock Lectures at the Cargese Summer School on QCD Perspectives on Hot and
  Dense Matter, Cargese, France, 6-18 Aug 2001.

\bibitem{Schukraft:2001vg}
J.~Schukraft,
\newblock Nucl. Phys. {\bf A698}, 287 (2002).

\bibitem{Adams:2003im}
STAR, J.~Adams {\em et~al.},
\newblock Phys. Rev. Lett. {\bf 91}, 072304 (2003), [nucl-ex/0306024].

\bibitem{Adler:2003ii}
PHENIX, S.~S. Adler {\em et~al.},
\newblock Phys. Rev. Lett. {\bf 91}, 072303 (2003), [nucl-ex/0306021].

\bibitem{Arsene:2004ux}
BRAHMS, I.~Arsene {\em et~al.},
\newblock nucl-ex/0403005.

\bibitem{Back:2003ns}
PHOBOS, B.~B. Back {\em et~al.},
\newblock Phys. Rev. Lett. {\bf 91}, 072302 (2003), [nucl-ex/0306025].

\bibitem{Baier:2003hr}
R.~Baier, A.~Kovner and U.~A. Wiedemann,
\newblock Phys. Rev. {\bf D68}, 054009 (2003), [hep-ph/0305265].

\bibitem{Kharzeev:2004bw}
D.~Kharzeev, E.~Levin and L.~McLerran,
\newblock hep-ph/0403271.

\bibitem{Kharzeev:2004yx}
D.~Kharzeev, Y.~V. Kovchegov and K.~Tuchin,
\newblock hep-ph/0405045.

\bibitem{Gyulassy:2003mc}
M.~Gyulassy, I.~Vitev, X.-N. Wang and B.-W. Zhang,
\newblock nucl-th/0302077.

\bibitem{Adler:2002tq}
STAR, C.~Adler {\em et~al.},
\newblock Phys. Rev. Lett. {\bf 90}, 082302 (2003), [nucl-ex/0210033].

\bibitem{Adler:2004zd}
PHENIX, S.~S. Adler {\em et~al.},
\newblock nucl-ex/0408007.

\bibitem{Braun-Munzinger:2003zd}
P.~Braun-Munzinger, K.~Redlich and J.~Stachel,
\newblock nucl-th/0304013.

\bibitem{Becattini:1995if}
F.~Becattini,
\newblock Z. Phys. {\bf C69}, 485 (1996).

\bibitem{Becattini:1997rv}
F.~Becattini and U.~W. Heinz,
\newblock Z. Phys. {\bf C76}, 269 (1997), [hep-ph/9702274].

\bibitem{Becattini:2001fg}
F.~Becattini and G.~Passaleva,
\newblock Eur. Phys. J. {\bf C23}, 551 (2002), [hep-ph/0110312].

\bibitem{Blanchard:2004du}
P.~Blanchard, S.~Fortunato and H.~Satz,
\newblock Eur. Phys. J. {\bf C34}, 361 (2004), [hep-ph/0401103].

\bibitem{Adams:2004bi}
STAR, J.~Adams {\em et~al.},
\newblock nucl-ex/0409033.

\bibitem{Huovinen:2001cy}
P.~Huovinen, P.~F. Kolb, U.~W. Heinz, P.~V. Ruuskanen and S.~A. Voloshin,
\newblock Phys. Lett. {\bf B503}, 58 (2001), [hep-ph/0101136].

\bibitem{Kolb:2002ve}
P.~F. Kolb and R.~Rapp,
\newblock Phys. Rev. {\bf C67}, 044903 (2003), [hep-ph/0210222].

\bibitem{Kolb:2004pi}
P.~F. Kolb,
\newblock nucl-th/0407066.

\bibitem{Kolb:2000sd}
P.~F. Kolb, J.~Sollfrank and U.~W. Heinz,
\newblock Phys. Rev. {\bf C62}, 054909 (2000), [hep-ph/0006129].

\bibitem{Retiere:2003kf}
F.~Retiere and M.~A. Lisa,
\newblock nucl-th/0312024.

\bibitem{Eskola:2002wx}
K.~J. Eskola, H.~Niemi, P.~V. Ruuskanen and S.~S. Rasanen,
\newblock Phys. Lett. {\bf B566}, 187 (2003), [hep-ph/0206230].

\bibitem{Magestro:2004du}
D.~Magestro,
\newblock nucl-ex/0408014.

\bibitem{Aarts:1998td}
G.~Aarts and J.~Smit,
\newblock Nucl. Phys. {\bf B555}, 355 (1999), [hep-ph/9812413].

\bibitem{Salle:2000hd}
M.~Salle, J.~Smit and J.~C. Vink,
\newblock Phys. Rev. {\bf D64}, 025016 (2001), [hep-ph/0012346].

\bibitem{Salle:2000jb}
M.~Salle, J.~Smit and J.~C. Vink,
\newblock Nucl. Phys. {\bf B625}, 495 (2002), [hep-ph/0012362].

\bibitem{Salle:2003ju}
M.~Salle,
\newblock Phys. Rev. {\bf D69}, 025005 (2004), [hep-ph/0307080].

\bibitem{Berges:2002wr}
J.~Berges, S.~Borsanyi and J.~Serreau,
\newblock Nucl. Phys. {\bf B660}, 51 (2003), [hep-ph/0212404].

\bibitem{Berges:2002cz}
J.~Berges and J.~Serreau,
\newblock Phys. Rev. Lett. {\bf 91}, 111601 (2003), [hep-ph/0208070].

\bibitem{Aarts:2002dj}
G.~Aarts, D.~Ahrensmeier, R.~Baier, J.~Berges and J.~Serreau,
\newblock Phys. Rev. {\bf D66}, 045008 (2002), [hep-ph/0201308].

\bibitem{Berges:2004yj}
J.~Berges,
\newblock hep-ph/0409233.

\bibitem{Berges:2004pu}
J.~Berges,
\newblock hep-ph/0401172.

\bibitem{Baier:2000sb}
R.~Baier, A.~H. Mueller, D.~Schiff and D.~T. Son,
\newblock Phys. Lett. {\bf B502}, 51 (2001), [hep-ph/0009237].

\bibitem{Arnold:2003rq}
P.~Arnold, J.~Lenaghan and G.~D. Moore,
\newblock JHEP {\bf 08}, 002 (2003), [hep-ph/0307325].

\bibitem{Elliott:1999uz}
D.~M. Elliott and D.~H. Rischke,
\newblock Nucl. Phys. {\bf A671}, 583 (2000), [nucl-th/9908004].

\bibitem{Arnold:2004ih}
P.~Arnold and J.~Lenaghan,
\newblock hep-ph/0408052.

\bibitem{Arnold:2004tf}
P.~Arnold,
\newblock hep-ph/0409002.

\bibitem{Arnold:2004ti}
P.~Arnold, J.~Lenaghan, G.~D. Moore and L.~G. Yaffe,
\newblock nucl-th/0409068.

\bibitem{Romatschke:2003ms}
P.~Romatschke and M.~Strickland,
\newblock Phys. Rev. {\bf D68}, 036004 (2003), [hep-ph/0304092].

\bibitem{Romatschke:2004jh}
P.~Romatschke and M.~Strickland,
\newblock hep-ph/0406188.

\bibitem{Romatschke:2004au}
P.~Romatschke and M.~Strickland,
\newblock hep-ph/0408275.

\bibitem{Romatschke:2004ma}
P.~Romatschke and M.~Strickland,
\newblock hep-ph/0408314.

\bibitem{Arnold:2000dr}
P.~Arnold, G.~D. Moore and L.~G. Yaffe,
\newblock JHEP {\bf 11}, 001 (2000), [hep-ph/0010177].

\bibitem{Arnold:2003zc}
P.~Arnold, G.~D. Moore and L.~G. Yaffe,
\newblock JHEP {\bf 05}, 051 (2003), [hep-ph/0302165].

\bibitem{Policastro:2001yc}
G.~Policastro, D.~T. Son and A.~O. Starinets,
\newblock Phys. Rev. Lett. {\bf 87}, 081601 (2001), [hep-th/0104066].

\bibitem{Wang:1991ht}
X.-N. Wang and M.~Gyulassy,
\newblock Phys. Rev. {\bf D44}, 3501 (1991).

\bibitem{Adcox:2001ry}
PHENIX, K.~Adcox {\em et~al.},
\newblock Phys. Rev. Lett. {\bf 87}, 052301 (2001), [nucl-ex/0104015].

\bibitem{Adler:2001yq}
STAR, C.~Adler {\em et~al.},
\newblock Phys. Rev. Lett. {\bf 87}, 112303 (2001), [nucl-ex/0106004].

\bibitem{Back:2001bq}
PHOBOS, B.~B. Back {\em et~al.},
\newblock Phys. Rev. Lett. {\bf 87}, 102303 (2001), [nucl-ex/0106006].

\bibitem{Bearden:2001xw}
BRAHMS, I.~G. Bearden {\em et~al.},
\newblock Phys. Lett. {\bf B523}, 227 (2001), [nucl-ex/0108016].

\bibitem{Bearden:2001qq}
BRAHMS, I.~G. Bearden {\em et~al.},
\newblock Phys. Rev. Lett. {\bf 88}, 202301 (2002), [nucl-ex/0112001].

\bibitem{Back:2002uc}
PHOBOS, B.~B. Back {\em et~al.},
\newblock Phys. Rev. {\bf C65}, 061901 (2002), [nucl-ex/0201005].

\bibitem{Adler:2004zn}
PHENIX, S.~S. Adler {\em et~al.},
\newblock nucl-ex/0409015.

\bibitem{Adams:2004cb}
STAR, J.~Adams {\em et~al.},
\newblock nucl-ex/0407003.

\bibitem{Heinz:2002rs}
U.~W. Heinz and S.~M.~H. Wong,
\newblock Phys. Rev. {\bf C66}, 014907 (2002), [hep-ph/0205058].

\bibitem{Eskola:1999fc}
K.~J. Eskola, K.~Kajantie, P.~V. Ruuskanen and K.~Tuominen,
\newblock Nucl. Phys. {\bf B570}, 379 (2000), [hep-ph/9909456].

\bibitem{Eskola:2001bf}
K.~J. Eskola, P.~V. Ruuskanen, S.~S. Rasanen and K.~Tuominen,
\newblock Nucl. Phys. {\bf A696}, 715 (2001), [hep-ph/0104010].

\bibitem{Hirano:2004rsb}
T.~Hirano and Y.~Nara,
\newblock nucl-th/0403029.

\bibitem{Hirano:2004rs}
T.~Hirano and Y.~Nara,
\newblock nucl-th/0404039.

\bibitem{Kharzeev:2000ph}
D.~Kharzeev and M.~Nardi,
\newblock Phys. Lett. {\bf B507}, 121 (2001), [nucl-th/0012025].

\bibitem{Khoze:1996dn}
V.~A. Khoze and W.~Ochs,
\newblock Int. J. Mod. Phys. {\bf A12}, 2949 (1997), [hep-ph/9701421].

\bibitem{Mueller:1999fp}
A.~H. Mueller,
\newblock Nucl. Phys. {\bf B572}, 227 (2000), [hep-ph/9906322].

\bibitem{Breitweg:1998dz}
ZEUS, J.~Breitweg {\em et~al.},
\newblock Eur. Phys. J. {\bf C7}, 609 (1999), [hep-ex/9809005].

\bibitem{Chekanov:2002pv}
ZEUS, S.~Chekanov {\em et~al.},
\newblock Phys. Rev. {\bf D67}, 012007 (2003), [hep-ex/0208023].

\bibitem{Gribov:1984tu}
L.~V. Gribov, E.~M. Levin and M.~G. Ryskin,
\newblock Phys. Rept. {\bf 100}, 1 (1983).

\bibitem{Mueller:1985wy}
A.~H. Mueller and J.-W. Qiu,
\newblock Nucl. Phys. {\bf B268}, 427 (1986).

\bibitem{Mueller:1999wm}
A.~H. Mueller,
\newblock Nucl. Phys. {\bf B558}, 285 (1999), [hep-ph/9904404].

\bibitem{Eskola:2002yc}
K.~J. Eskola, H.~Honkanen, V.~J. Kolhinen, J.-w. Qiu and C.~A. Salgado,
\newblock Nucl. Phys. {\bf B660}, 211 (2003), [hep-ph/0211239].

\bibitem{Eskola:2004ua}
K.~J. Eskola, V.~J. Kolhinen and R.~Vogt,
\newblock J. Phys. {\bf G30}, S1171 (2004), [hep-ph/0403130].

\bibitem{Golec-Biernat:1998js}
K.~Golec-Biernat and M.~Wusthoff,
\newblock Phys. Rev. {\bf D59}, 014017 (1999), [hep-ph/9807513].

\bibitem{Golec-Biernat:1999qd}
K.~Golec-Biernat and M.~Wusthoff,
\newblock Phys. Rev. {\bf D60}, 114023 (1999), [hep-ph/9903358].

\bibitem{Mueller:1993rr}
A.~H. Mueller,
\newblock Nucl. Phys. {\bf B415}, 373 (1994).

\bibitem{Mueller:1994gb}
A.~H. Mueller,
\newblock Nucl. Phys. {\bf B437}, 107 (1995), [hep-ph/9408245].

\bibitem{Mueller:1994jq}
A.~H. Mueller and B.~Patel,
\newblock Nucl. Phys. {\bf B425}, 471 (1994), [hep-ph/9403256].

\bibitem{Andersson:1990dp}
B.~Andersson, G.~Gustafson, A.~Nilsson and C.~Sjogren,
\newblock Z. Phys. {\bf C49}, 79 (1991).

\bibitem{Nikolaev:1990ja}
N.~N. Nikolaev and B.~G. Zakharov,
\newblock Z. Phys. {\bf C49}, 607 (1991).

\bibitem{McDermott:1999fa}
M.~McDermott, L.~Frankfurt, V.~Guzey and M.~Strikman,
\newblock Eur. Phys. J. {\bf C16}, 641 (2000), [hep-ph/9912547].

\bibitem{Brodsky:2002ue}
S.~J. Brodsky, P.~Hoyer, N.~Marchal, S.~Peigne and F.~Sannino,
\newblock Phys. Rev. {\bf D65}, 114025 (2002), [hep-ph/0104291].

\bibitem{Brodsky:2004hi}
S.~J. Brodsky, R.~Enberg, P.~Hoyer and G.~Ingelman,
\newblock hep-ph/0409119.

\bibitem{Derrick:1995ef}
ZEUS, M.~Derrick {\em et~al.},
\newblock Z. Phys. {\bf C69}, 607 (1996), [hep-ex/9510009].

\bibitem{Aid:1996au}
H1, S.~Aid {\em et~al.},
\newblock Nucl. Phys. {\bf B470}, 3 (1996), [hep-ex/9603004].

\bibitem{Derrick:1996hn}
ZEUS, M.~Derrick {\em et~al.},
\newblock Z. Phys. {\bf C72}, 399 (1996), [hep-ex/9607002].

\bibitem{Adloff:1997mf}
H1, C.~Adloff {\em et~al.},
\newblock Nucl. Phys. {\bf B497}, 3 (1997), [hep-ex/9703012].

\bibitem{Breitweg:1997hz}
ZEUS, J.~Breitweg {\em et~al.},
\newblock Phys. Lett. {\bf B407}, 432 (1997), [hep-ex/9707025].

\bibitem{Adloff:1997sc}
H1, C.~Adloff {\em et~al.},
\newblock Z. Phys. {\bf C76}, 613 (1997), [hep-ex/9708016].

\bibitem{Breitweg:1997aa}
ZEUS, J.~Breitweg {\em et~al.},
\newblock Eur. Phys. J. {\bf C1}, 81 (1998), [hep-ex/9709021].

\bibitem{Breitweg:1998gc}
ZEUS, J.~Breitweg {\em et~al.},
\newblock Eur. Phys. J. {\bf C6}, 43 (1999), [hep-ex/9807010].

\bibitem{Stasto:2000er}
A.~M. Stasto, K.~Golec-Biernat and J.~Kwiecinski,
\newblock Phys. Rev. Lett. {\bf 86}, 596 (2001), [hep-ph/0007192].

\bibitem{Adams:1995is}
E665, M.~R. Adams {\em et~al.},
\newblock Z. Phys. {\bf C67}, 403 (1995), [hep-ex/9505006].

\bibitem{Adams:1994ri}
E665, M.~R. Adams {\em et~al.},
\newblock Z. Phys. {\bf C65}, 225 (1995).

\bibitem{Arneodo:1996kd}
New Muon, M.~Arneodo {\em et~al.},
\newblock Nucl. Phys. {\bf B487}, 3 (1997), [hep-ex/9611022].

\bibitem{Wong:1970fu}
S.~K. Wong,
\newblock Nuovo Cim. {\bf A65}, 689 (1970).

\bibitem{Kovchegov:1996ty}
Y.~V. Kovchegov,
\newblock Phys. Rev. {\bf D54}, 5463 (1996), [hep-ph/9605446].

\bibitem{Kovchegov:1997pc}
Y.~V. Kovchegov,
\newblock Phys. Rev. {\bf D55}, 5445 (1997), [hep-ph/9701229].

\bibitem{Kovner:1995ja}
A.~Kovner, L.~D. McLerran and H.~Weigert,
\newblock Phys. Rev. {\bf D52}, 6231 (1995), [hep-ph/9502289].

\bibitem{Kovner:1995ts}
A.~Kovner, L.~D. McLerran and H.~Weigert,
\newblock Phys. Rev. {\bf D52}, 3809 (1995), [hep-ph/9505320].

\bibitem{Jalilian-Marian:1997xn}
J.~Jalilian-Marian, A.~Kovner, L.~D. McLerran and H.~Weigert,
\newblock Phys. Rev. {\bf D55}, 5414 (1997), [hep-ph/9606337].

\bibitem{Jalilian-Marian:2000ad}
J.~Jalilian-Marian, S.~Jeon and R.~Venugopalan,
\newblock Phys. Rev. {\bf D63}, 036004 (2001), [hep-ph/0003070].

\bibitem{Jeon:2004rk}
S.~Jeon and R.~Venugopalan,
\newblock hep-ph/0406169.

\bibitem{Lam:1999wu}
C.~S. Lam and G.~Mahlon,
\newblock Phys. Rev. {\bf D61}, 014005 (2000), [hep-ph/9907281].

\bibitem{Lam:2001ax}
C.~S. Lam and G.~Mahlon,
\newblock Phys. Rev. {\bf D64}, 016004 (2001), [hep-ph/0102337].

\bibitem{Krasnitz:2002ng}
A.~Krasnitz, Y.~Nara and R.~Venugopalan,
\newblock Phys. Lett. {\bf B554}, 21 (2003), [hep-ph/0204361].

\bibitem{Krasnitz:2002mn}
A.~Krasnitz, Y.~Nara and R.~Venugopalan,
\newblock Nucl. Phys. {\bf A717}, 268 (2003), [hep-ph/0209269].

\bibitem{Gyulassy:1997vt}
M.~Gyulassy and L.~D. McLerran,
\newblock Phys. Rev. {\bf C56}, 2219 (1997), [nucl-th/9704034].

\bibitem{Lam:2000nz}
C.~S. Lam and G.~Mahlon,
\newblock Phys. Rev. {\bf D62}, 114023 (2000), [hep-ph/0007133].

\bibitem{Jalilian-Marian:1997jx}
J.~Jalilian-Marian, A.~Kovner, A.~Leonidov and H.~Weigert,
\newblock Nucl. Phys. {\bf B504}, 415 (1997), [hep-ph/9701284].

\bibitem{Jalilian-Marian:1997gr}
J.~Jalilian-Marian, A.~Kovner, A.~Leonidov and H.~Weigert,
\newblock Phys. Rev. {\bf D59}, 014014 (1999), [hep-ph/9706377].

\bibitem{Jalilian-Marian:1997dw}
J.~Jalilian-Marian, A.~Kovner and H.~Weigert,
\newblock Phys. Rev. {\bf D59}, 014015 (1999), [hep-ph/9709432].

\bibitem{Weigert:2000gi}
H.~Weigert,
\newblock Nucl. Phys. {\bf A703}, 823 (2002), [hep-ph/0004044].

\bibitem{Iancu:2001ad}
E.~Iancu, A.~Leonidov and L.~D. McLerran,
\newblock Phys. Lett. {\bf B510}, 133 (2001), [hep-ph/0102009].

\bibitem{Mueller:2001uk}
A.~H. Mueller,
\newblock Phys. Lett. {\bf B523}, 243 (2001), [hep-ph/0110169].

\bibitem{Iancu:2001md}
E.~Iancu and L.~D. McLerran,
\newblock Phys. Lett. {\bf B510}, 145 (2001), [hep-ph/0103032].

\bibitem{Iancu:2002aq}
E.~Iancu, K.~Itakura and L.~McLerran,
\newblock Nucl. Phys. {\bf A724}, 181 (2003), [hep-ph/0212123].

\bibitem{Iancu:2002tr}
E.~Iancu, K.~Itakura and L.~McLerran,
\newblock Nucl. Phys. {\bf A708}, 327 (2002), [hep-ph/0203137].

\bibitem{Blaizot:2002xy}
J.-P. Blaizot, E.~Iancu and H.~Weigert,
\newblock Nucl. Phys. {\bf A713}, 441 (2003), [hep-ph/0206279].

\bibitem{Rummukainen:2003ns}
K.~Rummukainen and H.~Weigert,
\newblock Nucl. Phys. {\bf A739}, 183 (2004), [hep-ph/0309306].

\bibitem{Iancu:2002xk}
E.~Iancu, A.~Leonidov and L.~McLerran,
\newblock hep-ph/0202270.

\bibitem{Iancu:2003xm}
E.~Iancu and R.~Venugopalan,
\newblock hep-ph/0303204.

\bibitem{Balitsky:1995ub}
I.~Balitsky,
\newblock Nucl. Phys. {\bf B463}, 99 (1996), [hep-ph/9509348].

\bibitem{Buchmuller:1996xw}
W.~Buchmuller, M.~F. McDermott and A.~Hebecker,
\newblock Nucl. Phys. {\bf B487}, 283 (1997), [hep-ph/9607290].

\bibitem{Buchmuller:1998jv}
W.~Buchmuller, T.~Gehrmann and A.~Hebecker,
\newblock Nucl. Phys. {\bf B537}, 477 (1999), [hep-ph/9808454].

\bibitem{Mueller:2001fv}
A.~H. Mueller,
\newblock hep-ph/0111244.

\bibitem{Gavai:1996vu}
R.~V. Gavai and R.~Venugopalan,
\newblock Phys. Rev. {\bf D54}, 5795 (1996), [hep-ph/9605327].

\bibitem{Kovchegov:1999yj}
Y.~V. Kovchegov,
\newblock Phys. Rev. {\bf D60}, 034008 (1999), [hep-ph/9901281].

\bibitem{Lam:2002mg}
C.~S. Lam, G.~Mahlon and W.~Zhu,
\newblock Phys. Rev. {\bf D66}, 074005 (2002), [hep-ph/0207058].

\bibitem{Kovner:2000pt}
A.~Kovner, J.~G. Milhano and H.~Weigert,
\newblock Phys. Rev. {\bf D62}, 114005 (2000), [hep-ph/0004014].

\bibitem{Kovchegov:2000hz}
Y.~V. Kovchegov,
\newblock Nucl. Phys. {\bf A692}, 557 (2001), [hep-ph/0011252].

\bibitem{Armesto:1996kt}
N.~Armesto, M.~A. Braun, E.~G. Ferreiro and C.~Pajares,
\newblock Phys. Rev. Lett. {\bf 77}, 3736 (1996), [hep-ph/9607239].

\bibitem{Nardi:1998qb}
M.~Nardi and H.~Satz,
\newblock Phys. Lett. {\bf B442}, 14 (1998), [hep-ph/9805247].

\bibitem{Satz:2002qj}
H.~Satz,
\newblock hep-ph/0212046.

\bibitem{Digal:2002bm}
S.~Digal, S.~Fortunato, P.~Petreczky and H.~Satz,
\newblock Phys. Lett. {\bf B549}, 101 (2002), [hep-ph/0207264].

\bibitem{Digal:2003sg}
S.~Digal, S.~Fortunato and H.~Satz,
\newblock Eur. Phys. J. {\bf C32}, 547 (2004), [hep-ph/0310354].

\bibitem{Dumitru:2001ux}
A.~Dumitru and L.~D. McLerran,
\newblock Nucl. Phys. {\bf A700}, 492 (2002), [hep-ph/0105268].

\bibitem{Kovchegov:1997ke}
Y.~V. Kovchegov and D.~H. Rischke,
\newblock Phys. Rev. {\bf C56}, 1084 (1997), [hep-ph/9704201].

\bibitem{Kovchegov:1998bi}
Y.~V. Kovchegov and A.~H. Mueller,
\newblock Nucl. Phys. {\bf B529}, 451 (1998), [hep-ph/9802440].

\bibitem{Makhlin:1996dt}
A.~Makhlin,
\newblock hep-ph/9608259.

\bibitem{Makhlin:1996dr}
A.~Makhlin,
\newblock hep-ph/9608261.

\bibitem{Makhlin:1998zi}
A.~Makhlin and E.~Surdutovich,
\newblock Phys. Rev. {\bf C58}, 389 (1998), [hep-ph/9803364].

\bibitem{Makhlin:2000nw}
A.~Makhlin,
\newblock Phys. Rev. {\bf C63}, 044902 (2001), [hep-ph/0007300].

\bibitem{Makhlin:2000nx}
A.~Makhlin,
\newblock Phys. Rev. {\bf C63}, 044903 (2001), [hep-ph/0007301].

\bibitem{Makhlin:2000ny}
A.~Makhlin and E.~Surdutovich,
\newblock Phys. Rev. {\bf C63}, 044904 (2001), [hep-ph/0007302].

\bibitem{Makhlin:2000xf}
A.~Makhlin,
\newblock hep-ph/0009067.

\bibitem{Gunion:1981qs}
J.~F. Gunion and G.~Bertsch,
\newblock Phys. Rev. {\bf D25}, 746 (1982).

\bibitem{Eskola:1996ce}
K.~J. Eskola and K.~Kajantie,
\newblock Z. Phys. {\bf C75}, 515 (1997), [nucl-th/9610015].

\bibitem{Gelis:2003vh}
F.~Gelis and R.~Venugopalan,
\newblock Phys. Rev. {\bf D69}, 014019 (2004), [hep-ph/0310090].

\bibitem{Shuryak:2002qz}
E.~Shuryak and I.~Zahed,
\newblock Phys. Rev. {\bf D67}, 014006 (2003), [hep-ph/0206022].

\bibitem{Kajantie:1985jh}
K.~Kajantie and T.~Matsui,
\newblock Phys. Lett. {\bf B164}, 373 (1985).

\bibitem{Gatoff:1987uf}
G.~Gatoff, A.~K. Kerman and T.~Matsui,
\newblock Phys. Rev. {\bf D36}, 114 (1987).

\bibitem{Kluger:1992gb}
Y.~Kluger, J.~M. Eisenberg, B.~Svetitsky, F.~Cooper and E.~Mottola,
\newblock Phys. Rev. {\bf D45}, 4659 (1992).

\bibitem{Cooper:1993hw}
F.~Cooper, J.~M. Eisenberg, Y.~Kluger, E.~Mottola and B.~Svetitsky,
\newblock Phys. Rev. {\bf D48}, 190 (1993), [hep-ph/9212206].

\bibitem{Dietrich:2003qf}
D.~D. Dietrich,
\newblock Phys. Rev. {\bf D68}, 105005 (2003), [hep-th/0302229].

\bibitem{Dietrich:2004eb}
D.~D. Dietrich,
\newblock hep-th/0402026.

\bibitem{Baltz:1998zb}
A.~J. Baltz and L.~D. McLerran,
\newblock Phys. Rev. {\bf C58}, 1679 (1998), [nucl-th/9804042].

\bibitem{Baltz:2001dp}
A.~J. Baltz, F.~Gelis, L.~D. McLerran and A.~Peshier,
\newblock Nucl. Phys. {\bf A695}, 395 (2001), [nucl-th/0101024].

\bibitem{Birrell:1982ix}
N.~D. Birrell and P.~C.~W. Davies,
\newblock {\em Quantum fields in curved space} (Cambridge University Press,
  1982).

\bibitem{Krasnitz:2000gz}
A.~Krasnitz and R.~Venugopalan,
\newblock Phys. Rev. Lett. {\bf 86}, 1717 (2001), [hep-ph/0007108].

\bibitem{Krasnitz:1999wc}
A.~Krasnitz and R.~Venugopalan,
\newblock Phys. Rev. Lett. {\bf 84}, 4309 (2000), [hep-ph/9909203].

\bibitem{Krasnitz:2001qu}
A.~Krasnitz, Y.~Nara and R.~Venugopalan,
\newblock Phys. Rev. Lett. {\bf 87}, 192302 (2001), [hep-ph/0108092].

\bibitem{Krasnitz:2003jw}
A.~Krasnitz, Y.~Nara and R.~Venugopalan,
\newblock Nucl. Phys. {\bf A727}, 427 (2003), [hep-ph/0305112].

\bibitem{Jalilian-Marian:2003mf}
J.~Jalilian-Marian, Y.~Nara and R.~Venugopalan,
\newblock Phys. Lett. {\bf B577}, 54 (2003), [nucl-th/0307022].

\bibitem{Kogut:1974ag}
J.~B. Kogut and L.~Susskind,
\newblock Phys. Rev. {\bf D11}, 395 (1975).

\bibitem{Kogut:1982ds}
J.~B. Kogut,
\newblock Rev. Mod. Phys. {\bf 55}, 775 (1983).

\bibitem{Krasnitz:1995xi}
A.~Krasnitz,
\newblock Nucl. Phys. {\bf B455}, 320 (1995), [hep-lat/9507025].

\bibitem{Ambjorn:1995xm}
J.~Ambjorn and A.~Krasnitz,
\newblock Phys. Lett. {\bf B362}, 97 (1995), [hep-ph/9508202].

\bibitem{Ambjorn:1997jz}
J.~Ambjorn and A.~Krasnitz,
\newblock Nucl. Phys. {\bf B506}, 387 (1997), [hep-ph/9705380].

\bibitem{Bodeker:1999gx}
D.~Bodeker, G.~D. Moore and K.~Rummukainen,
\newblock Phys. Rev. {\bf D61}, 056003 (2000), [hep-ph/9907545].

\bibitem{numrec}
W.~H. Press, S.~A. Teukolsky, W.~T. Vetterling and B.~P. Flannery,
\newblock {\em Numerical Recipes in C}, 2nd ed. (Cambridge University Press,
  1992).

\bibitem{Tranberg:2003gi}
A.~Tranberg and J.~Smit,
\newblock JHEP {\bf 11}, 016 (2003), [hep-ph/0310342].

\bibitem{Bearden:2004yx}
BRAHMS, I.~G. Bearden,
\newblock nucl-ex/0403050.

\bibitem{Adams:2003xp}
STAR, J.~Adams {\em et~al.},
\newblock Phys. Rev. Lett. {\bf 92}, 112301 (2004), [nucl-ex/0310004].

\bibitem{Eskola:2002qz}
K.~J. Eskola, K.~Kajantie, P.~V. Ruuskanen and K.~Tuominen,
\newblock Phys. Lett. {\bf B543}, 208 (2002), [hep-ph/0204034].

\bibitem{Tuominen:2002sq}
K.~Tuominen,
\newblock Nucl. Phys. {\bf A715}, 809 (2003), [hep-ph/0209102].

\bibitem{Eskola:1997hz}
K.~J. Eskola, K.~Kajantie and P.~V. Ruuskanen,
\newblock Eur. Phys. J. {\bf C1}, 627 (1998), [nucl-th/9705015].

\bibitem{Kovchegov:2002nf}
Y.~V. Kovchegov and K.~L. Tuchin,
\newblock Nucl. Phys. {\bf A708}, 413 (2002), [hep-ph/0203213].

\bibitem{weinberg2}
S.~Weinberg,
\newblock {\em The Quantum Theory of Fields, Vol. 2: Modern Applications}
  (Cambridge University Press, 1996).

\bibitem{Wilson:1974sk}
K.~G. Wilson,
\newblock Phys. Rev. {\bf D10}, 2445 (1974).

\end{thebibliography}

\end{document}